\documentclass{aa}  

\usepackage{graphicx}
\usepackage{txfonts}
\usepackage[colorlinks=true, allcolors=blue]{hyperref}

\usepackage{natbib}
\bibpunct{(}{)}{;}{a}{}{,}

\newcommand{\ergcm}[1]{$\times 10^{#1}$ erg cm$^{-2}$ s$^{-1}$}

\newcommand{\uergcm}{erg cm$^{-2}$ s$^{-1}$}
\newcommand{\ergs}[1]{$\times 10^{#1}$ erg s$^{-1}$}

\newcommand{\uergs}{erg s$^{-1}$}
\newcommand{\hcm}[1]{$\times 10^{#1}$ cm$^{-2}$}

\newcommand{\uhcm}{cm$^{-2}$}
\newcommand{\expo}[1]{$\times 10^{#1}$}
\newcommand{\oexpo}[1]{$10^{#1}$}

\newcommand{\nh}{N$_{\rm H}$\xspace}

\newcommand{\cts}{cts s$^{-1}$\xspace}

\newcommand{\Hone}{\ion{H}{i}\xspace}
\newcommand{\Hmol}{H$_2$\xspace}

\newcommand{\ltsima}{$\buildrel < \over \sim$}
\newcommand{\lsim}{\lower.5ex\hbox{\ltsima}}
\newcommand{\msun}{M$_{\odot}$}

\newcommand{\Mdot}{$\dot{\rm M}$\xspace}
\newcommand{\rahour}{\hbox{\ensuremath{^{\rm h}}}}
\newcommand{\ramin}{\hbox{\ensuremath{^{\rm m}}}}

\newcommand{\xspec}{\texttt{XSPEC}\xspace}
\newcommand{\eSASS}{\texttt{eSASS}\xspace}
\newcommand{\decam}{{DECam}\xspace}

\newcommand{\swift}{{\it Swift}\xspace}
\newcommand{\xmm}{{\it XMM-Newton}\xspace}

\newcommand{\nicer}{{\it NICER}\xspace}
\newcommand{\srg}{{\it SRG}\xspace}
\newcommand{\ero}{\mbox{eROSITA}\xspace}
\newcommand{\esrc}{\mbox{eRASSt\,J040515.6$-$745202}\xspace}

\graphicspath{{figs/}}

\begin{document} 

\title{\esrc, an X-ray burster in the Magellanic Bridge}

\author{F. Haberl\inst{\ref{mpe}} \and
        G. Vasilopoulos\inst{\ref{oas}} \and
        C. Maitra\inst{\ref{mpe}} \and
        F. Valdes\inst{\ref{noir}} \and
        D. Lang\inst{\ref{pitp}} \and
        V. Doroshenko\inst{\ref{iaat}} \and
        L. Ducci\inst{\ref{iaat}} \and
        I. Kreykenbohm\inst{\ref{uerl}} \and
        A. Rau\inst{\ref{mpe}} \and
        P. Weber\inst{\ref{uerl}} \and
        J. Wilms\inst{\ref{uerl}} \and
        P. Maggi\inst{\ref{oas}} \and
        C.~D.~Bailyn\inst{\ref{yale}} \and
        G.~K.~Jaisawal\inst{\ref{nsid}} \and
        P. S. Ray\inst{\ref{nrl}} \and
        H. Treiber\inst{\ref{amh}}
       } 

\titlerunning{An X-ray burster in the Magellanic Bridge}
\authorrunning{Haberl et al.}

\institute{
Max-Planck-Institut f{\"u}r extraterrestrische Physik, Gie{\ss}enbachstra{\ss}e 1, D-85748 Garching, Germany\label{mpe}, \email{fwh@mpe.mpg.de}
\and
Universit\'e de Strasbourg, CNRS, Observatoire astronomique de Strasbourg, UMR 7550, F-67000 Strasbourg, France\label{oas}
\and
NSF’s National Optical/Infrared Research Laboratory (NOIRLab), 950 N. Cherry Ave, Tucson, AZ 85732, USA\label{noir}
\and
Perimeter Institute for Theoretical Physics, 31 Caroline Street North, Waterloo, ON  N2L 2Y5, Canada\label{pitp}
\and
Institut f{\"u}r Astronomie und Astrophysik, Sand 1, D-72076 T{\"u}bingen, Germany\label{iaat}
\and
Remeis Observatory and ECAP, Universit{\"a}t Erlangen-N{\"u}rnberg, Sternwartstra{\ss}e 7, D-96049 Bamberg, Germany\label{uerl}
\and
Department of Astronomy, Yale University, PO Box 208101, New Haven, CT 06520-8101, USA\label{yale}
\and
DTU Space, Technical University of Denmark, Elektrovej 327-328, DK-2800 Lyngby, Denmark\label{nsid}
\and
U.S. Naval Research Laboratory, Washington, DC 20375, USA\label{nrl}
\and
Department of Physics and Astronomy, Amherst College, C025 New Science Center, 25 East Dr., Amherst, MA 01002-5000, USA\label{amh}
}

\date{Received 20 September 2022 / Accepted 4 November 2022}

\abstract
   {During the third all-sky survey (eRASS3), \ero, the soft X-ray instrument aboard {\it Spectrum-Roentgen-Gamma}, detected a new hard X-ray transient, \esrc, in the direction of the Magellanic Bridge.} 
   {We arranged follow-up observations and searched for archival data to reveal the nature of the transient.}
   {Using X-ray observations with \xmm, \nicer, and \swift, we investigated the temporal and spectral behaviour of the source for over about 10\,days.}
   {The X-ray light curve obtained from the \xmm observation with an  $\sim$28\,ks exposure revealed a type-I X-ray burst with a peak bolometric luminosity of at least 1.4\ergs{37}. 
   The burst energetics are consistent with a location of the burster at the distance of the Magellanic Bridge.  
   The relatively long exponential decay time of the burst of $\sim$70\,s indicates that it ignited in a H-rich environment.
   The non-detection of the source during the other \ero surveys, twelve and six months before and six months after eRASS3, 
   suggests that the burst was discovered during a moderate outburst which reached 2.6\ergs{36} in persistent emission. 
   During the \nicer observations, the source showed alternating flux states with the high level at a similar brightness as during the \xmm observation. 
   This behaviour is likely caused by dips as also seen during the last hour of the \xmm observation. Evidence for a recurrence of the dips with a period of $\sim$21.8\,hr suggests \esrc is a low-mass X-ray binary (LMXB) system with an accretion disk seen nearly edge on.
   We identify a multi-wavelength counterpart to the X-ray source in UVW1 and g, r, i, and z images obtained by the optical/UV monitor on \xmm 
   and the Dark Energy Camera at the Cerro Tololo Inter-American Observatory. 
   The spectral energy distribution is consistent with radiation from an accretion disk which dominates the UV and from a cool late-type star detected in the optical to infrared wavelengths.}
   {After the discovery of X-ray bursts in M\,31, the Magellanic Bridge is only the second location outside of the Milky Way where an X-ray burster was found. The burst uniquely identifies \esrc as an LMXB system with a neutron star. Its location in the Magellanic Bridge confirms the existence of an older stellar population which is expected if the bridge was formed by tidal interactions between the Magellanic Clouds, which stripped gas and stars from the clouds.}


\maketitle   


\section{Introduction}
\label{sec:intro}

Type-I X-ray bursts are observed from low-mass X-ray binaries (LMXBs) and they uniquely identify the compact object in the system as a neutron star. The bursts occur when a sufficiently large amount of matter rich in H and He, which is accreted from the companion star, is gathered on the surface of the neutron star to ignite a thermonuclear runaway.
Depending on the accretion rate, most bursts
repeat on timescales of hours to days and typically last from $\sim$10\,s to minutes.
A few bursts were observed with much longer durations between several hours and a day.
These very rare superbursts with a long recurrence time are thought to burn carbon, rather than the H-He fuel for short bursts 
\citep[for reviews, see][]{2021ASSL..461..209G,2013PrPNP..69..225P,2006csxs.book..113S,1993SSRv...62..223L}. Temporal and spectral properties of a  large sample of more than 7000 bursts from 85 bursting sources are described in \citet{2020ApJS..249...32G}. 

The first type-I X-ray bursts outside of the Milky Way were detected in the Local Group galaxy M\,31 using \xmm data \citep{2005A&A...430L..45P}.
The authors found two bursts, which were significantly detected in at least two of the EPIC instruments \citep{2001A&A...365L..18S,2001A&A...365L..27T}.
In a systematic search for fast transients in \xmm EPIC-pn observations, \citet{2020A&A...640A.124P} reported the identification of four bursts in the direction of M\,31. The first burst in their Table\,6 is identical to the second burst reported by \citet{2005A&A...430L..45P}\footnote{This was probably not recognised because a wrong observation ID was given by \citet{2005A&A...430L..45P} in the caption of their Fig.\,4.}. The other three bursts are new, thus increasing the number of X-ray bursters known in M\,31 to five.

In the course of our X-ray source population studies of the Large and Small Magellanic Cloud (LMC and SMC) with \ero, we also monitored the Magellanic Bridge between the two satellite galaxies of the Milky Way. Together with the Magellanic Clouds, the Magellanic Bridge is part of a gaseous structure around the Magellanic Clouds spanning 200 degrees on the sky in total, also including a leading arm and a trailing stream \citep{2010ApJ...723.1618N}.
The Magellanic Bridge consists of neutral gas and a young stellar component \citep{1990AJ.....99..191I}. Moreover, \citet{2013A&A...551A..78B} found evidence for an older stellar population. Such a population is expected if the bridge was formed by tidal interactions between the LMC and SMC, which stripped gas and stars from the clouds \citep{1985PASA....6..104M}.

During the third 6\,month long all-sky survey (eRASS3), the \ero instrument \citep{2021A&A...647A...1P} 
on board the Russian/German {\it Spektrum-Roentgen-Gamma} (\srg) mission 
\citep{2021A&A...656A.132S} 
discovered a new hard X-ray transient, designated \esrc \citep{2021ATel14646....1R}.
The source is located in the direction of the Magellanic Bridge, was not detected in eRASS1 nor eRASS2, and had faded again below the detection limit in eRASS4. 
Here, we report on X-ray and UV follow-up observations of \esrc with \xmm, \swift, and \nicer.  An optical identification is provided from Dark Energy Camera (\decam) archival exposures.

\section{X-ray observations}
\label{sec:xobs}

\subsection{\ero}
\label{sec:eroobs}

After its launch in July 2019, \ero started to scan the sky in December 2019 and completed four full sky surveys (eRASS1 to eRASS4) by December 2021. \ero consists of seven co-aligned Wolter type-I telescopes, each equipped with a camera based on a charge coupled device (CCD) of pn type with an integration time of 50\,ms \citep{2021A&A...647A...1P}.

\begin{figure}
\centering
  \resizebox{0.85\hsize}{!}{\includegraphics{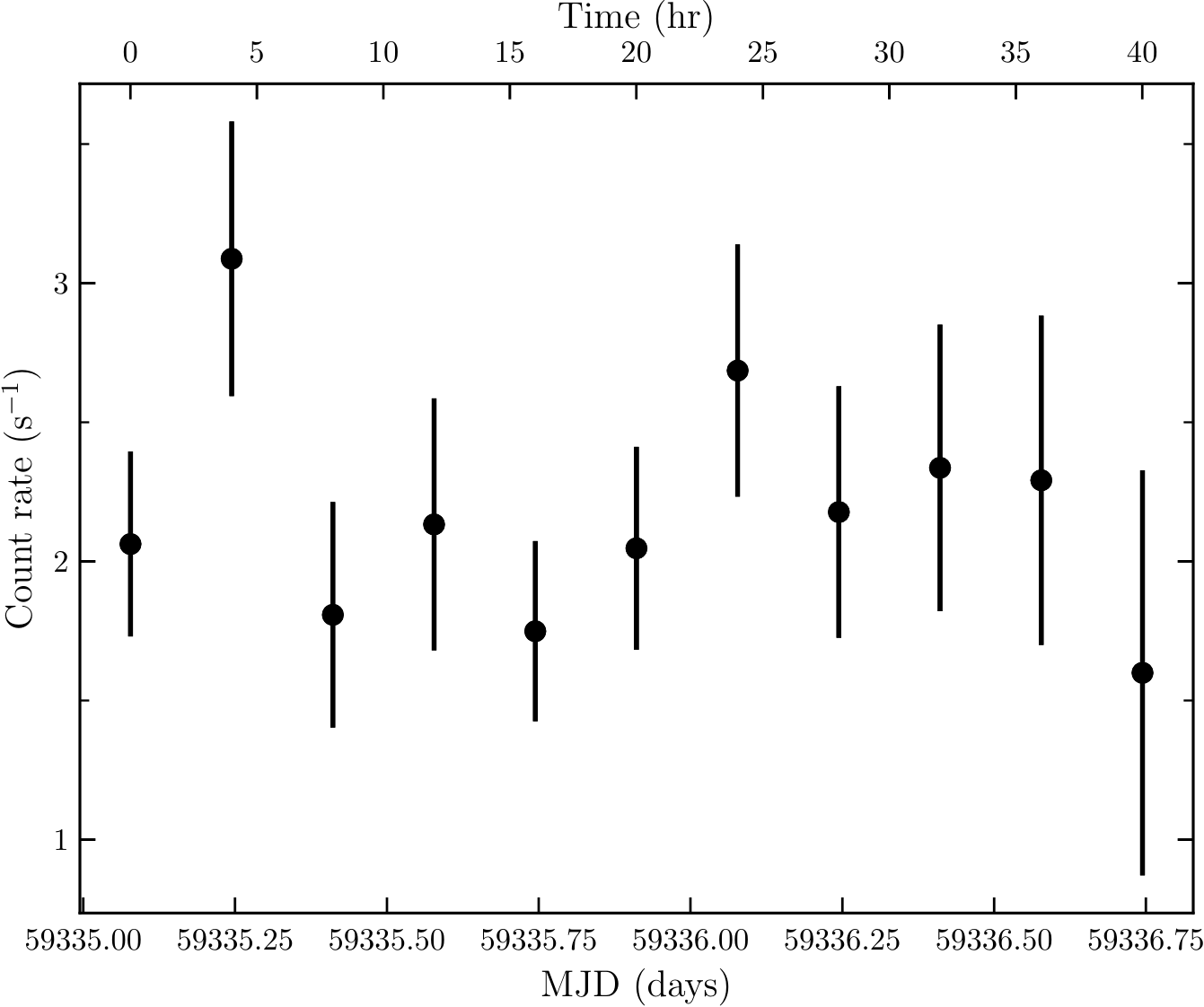}}
  \caption{Background-subtracted \ero light curve of \esrc during eRASS3 in the energy band 0.2--8.0\,keV. Each data point represents one scan with a typical duration of 40\,s
  and combines data from all seven cameras. The count rates were corrected for vignetting and point-spread function losses. The time in hours starts at the beginning of the first scan.
  }
  \label{fig:eroLC}
\end{figure}

\ero scanned \esrc eleven times during eRASS3 between 2021 May 1, 01:51 (MJD 59335.07738) and 2021 May 2, 17:52 (MJD 59336.74441; see Fig.\,\ref{fig:eroLC}).
For the data analysis, we used the \ero Standard Analysis Software System 
\citep[\eSASS version {\tt eSASSusers\_211214\_0\_3};][]{2022A&A...661A...1B}. 
To extract source and background events for light curves and spectra, we used the \eSASS task \texttt{srctool}  
\citep[see e.g. ][]{2021A&A...647A...8M,2022A&A...661A..25H}. 
We used circular regions with radii of 90\arcsec\ and 120\arcsec\ around the position of the source and a nearby source-free region, respectively, and selected all valid pixel patterns (PATTERN=15). For the light curves, we combined the data from all cameras (telescope modules TM 1--7) and applied a cut in the fractional exposure of 0.15 (FRACEXP$>$0.15) to exclude data from the edge of the detectors. 
We created a combined spectrum from the data of TMs 1--4 \& 6, the five cameras with an on-chip optical blocking filter, with an average exposure time of 294\,s. TM5 and TM7 suffer from a light leak \citep{2021A&A...647A...1P} and no reliable energy calibration is available yet.

\subsection{\swift}
\label{sec:swiftobs}
Following the \ero discovery of \esrc \citep{2021ATel14646....1R}, we initiated follow-up target of opportunity observations with the Neil Gehrels Swift Observatory \citep{2004ApJ...611.1005G}.
We used the data obtained from the X-ray telescope (XRT) in photon counting mode with a time resolution of 2.5\,s.
The observations were performed on 2021 May 12 and a total XRT exposure time of 1682\,s was obtained.
The XRT data were analysed using standard procedures \citep{2007A&A...469..379E,2009MNRAS.397.1177E}. 

\subsection{\xmm}
\label{sec:xmmobs}

To investigate the nature of \esrc, we triggered one of our \xmm anticipated target of opportunity observations (PI FH) to follow up on new hard transient sources in the Magellanic system.
For this purpose, we utilised data from the European Photon Imaging Camera (EPIC), sensitive to the 0.15--12\,keV band. Two of the three \xmm telescopes are equipped with Metal Oxide Semi-conductor (MOS) CCD arrays \citep{2001A&A...365L..27T} and the third with a pn-CCD \citep[a predecessor of the \ero CCDs;][]{2001A&A...365L..18S}.
The observation was performed on 2021 May 14 (observation ID 0860800401, start MJD = 59348.0497). 
We used the cameras with medium optical blocking filters, EPIC-pn in large-window readout mode (48\,ms time resolution), and EPIC-MOS in full-frame mode (2.6\,s).
We obtained net exposure times of 25.7\,ks and 27.9\,ks for EPIC-pn and EPIC-MOS, respectively, after we removed intervals of high background flaring activity, which occurred only at the end of the observation.

\xmm/EPIC data were processed using the \xmm data analysis software SAS, version 19.1.0\footnote{Science Analysis Software (SAS):\\ \url{http://xmm.esac.esa.int/sas/}}. 
We used the SAS task \texttt{evselect} for event extraction from circular regions around the source (45\arcsec\ radius) and nearby background (60\arcsec\ radius). Events with \texttt{PATTERN} 1--4 for EPIC-pn \citep{2001A&A...365L..18S} and \texttt{PATTERN} 1--12 for EPIC-MOS \citep{2001A&A...365L..27T} were selected. 
We created light curves in the energy bands 0.2--2.2\,keV, 2.2--8.0\,keV, and 0.2--8.0\,keV, applying the standard filtering with flags \texttt{\#XMMEA\_EP and \#XMMEA\_EM} for EPIC-pn and EPIC-MOS, respectively. 
For spectra we used the conservative FLAG=0 filtering and created response files using the SAS tasks \texttt{arfgen} and \texttt{rmfgen}.
As the best X-ray source position, we used the one derived by the \xmm pipeline of 
$\alpha_\mathrm{J2000.0}= 04\rahour\,05\ramin\,14\fs96$ and $\delta_\mathrm{J2000.0} = -74\degr\,52\arcmin\,01\farcs8$ with a $1\sigma$ statistical uncertainty of $0\farcs04$ and a remaining systematic error of $0\farcs91$ after astrometric correction.

\subsection{\nicer}
\label{sec:nicer}

The \nicer X-ray Timing Instrument \citep[XTI,][]{2012SPIE.8443E..13G,2016SPIE.9905E..1HG} is a non-imaging, soft X-ray telescope on board the \textit{International Space Station} (ISS). 
The XTI consists of an array of 56 co-aligned concentrator optics, each associated with a silicon drift detector \citep{2012SPIE.8453E..18P}, operating in the 0.2--12 keV band. 
The XTI provides high time resolution ($\sim$100 ns) and spectral  resolution of $\sim$85\,eV at 1\,keV. It has a field of view of $\sim$30 arcmin$^2$ in the sky and an effective area of ${\sim}1900\,\mathrm{cm}^2$ at 1.5\,keV (with 52 active detectors).

For the current study, we used \nicer data obtained between MJD 59339 and 59346 (i.e. before the \xmm observation) which were grouped into six observation intervals (observation IDs  4570010101--6).
Data were analysed with {\tt HEASOFT} version 6.29c,
\nicer DAS version 2020-04-23\_V007a, and the calibration database (CALDB) version 20200722 (redistribution matrix and ancillary response files). 
For event selection and background filtering, we used standard screening criteria and intervals with magnetic cut-off rigidity between 1--15 GeV/c.
Since \nicer is not an imaging instrument, the X-ray background must be calculated from a database of data collected from empty background fields. 
The `space weather' method (Gendreau et al., in prep.) contains functions which can be used to create the background spectrum based on estimates of the space weather environment.  
On the other hand, the {\tt 3C50} tool \citep{2022AJ....163..130R} uses a number of background proxies from each \nicer observation to define the basis states of the background database. 
We opted to use the latter since, for observations in the direction of the Magellanic Clouds, it has been shown to produce better estimates of the background \citep{2021MNRAS.503.6187T}. 
After filtering the data, we were left with an exposure of $\sim$12.8\,ks.

\section{X-ray data analysis}
\label{sec:xanalysis}

\begin{figure}
\centering
  \resizebox{0.92\hsize}{!}{\includegraphics{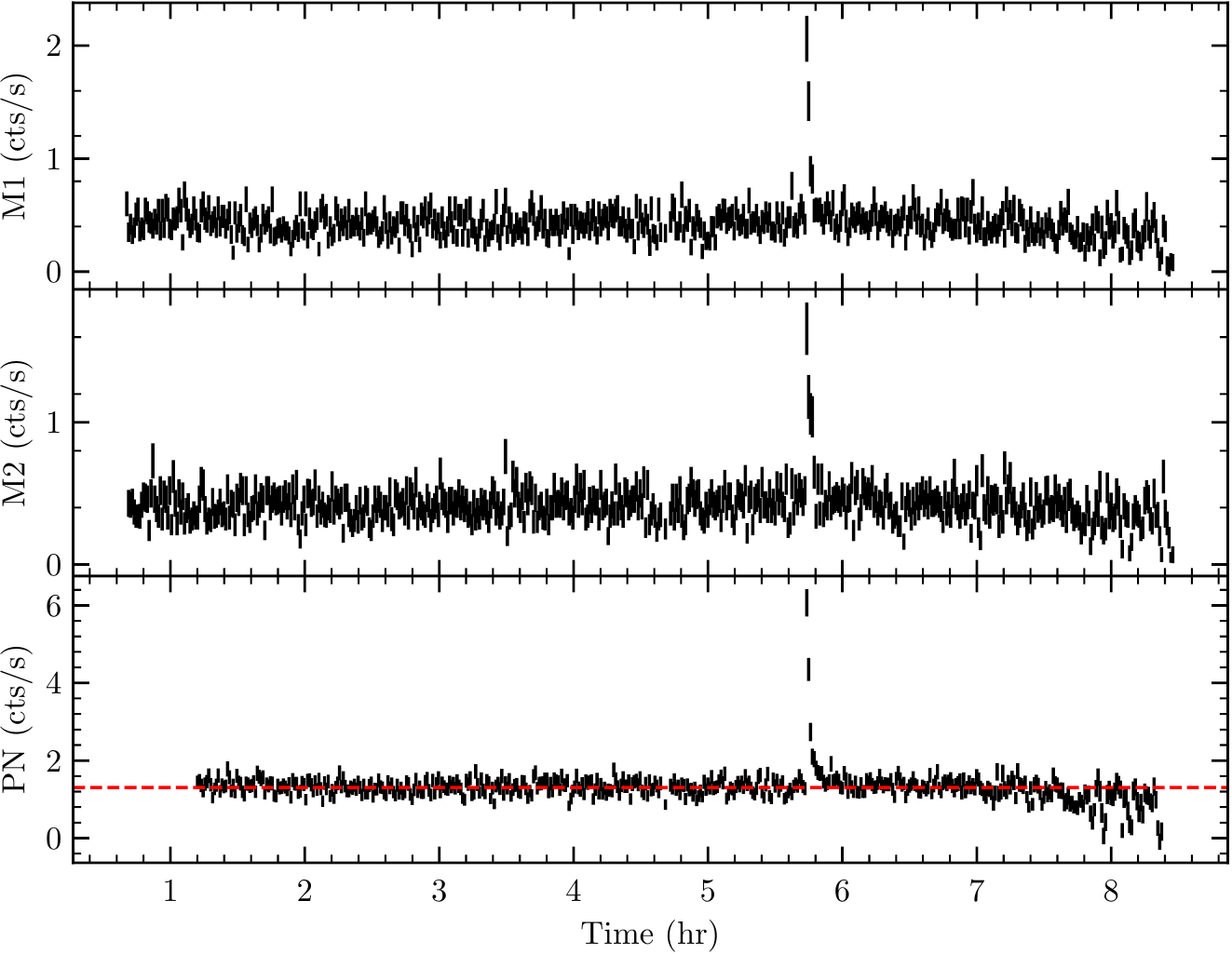}}
  \resizebox{0.92\hsize}{!}{\includegraphics{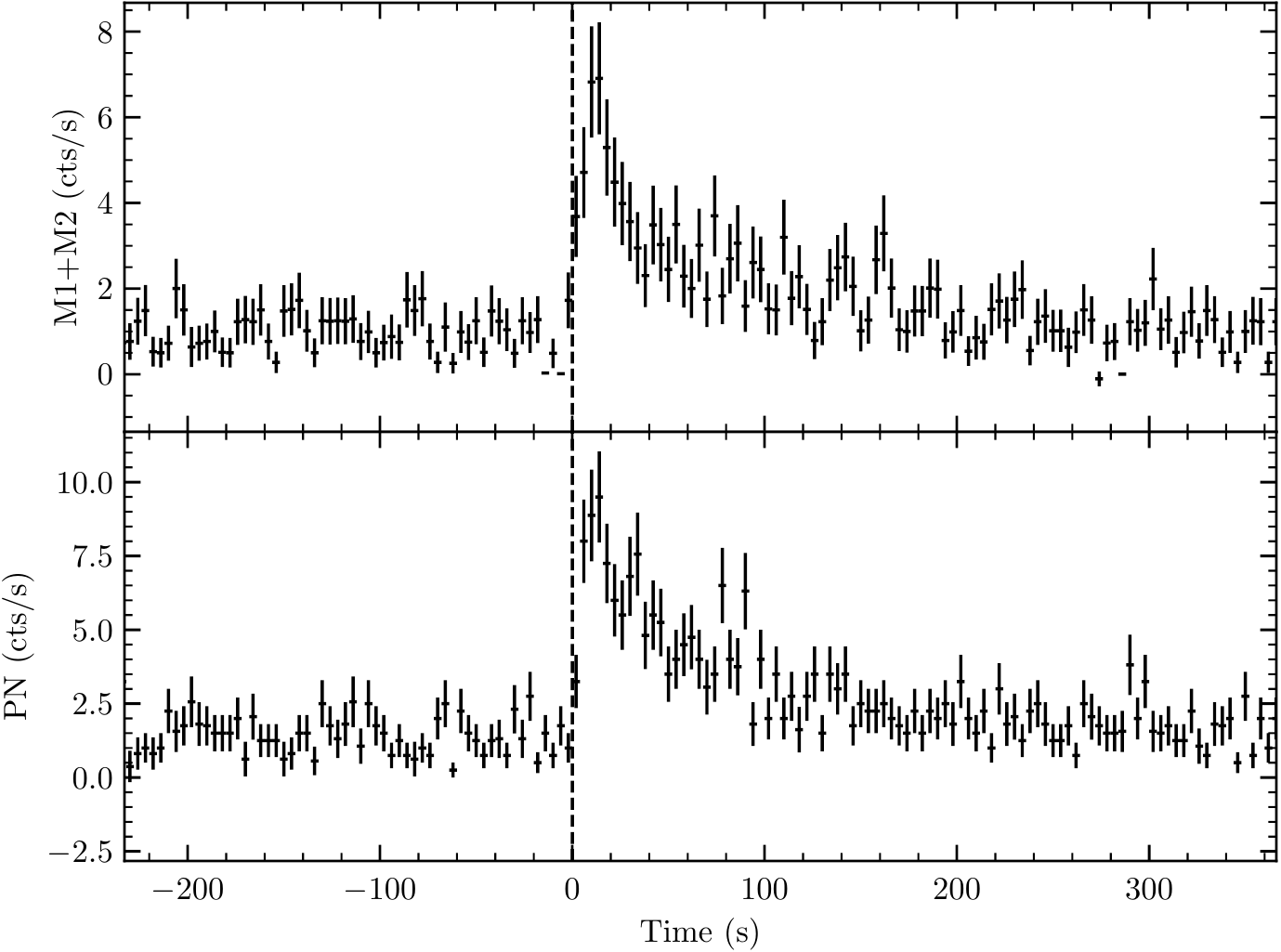}}
  \caption{Top: Background-subtracted EPIC-MOS1, MOS2, and pn light curves of \esrc in the energy band 0.2--8.0\,keV with time bins of 50\,s. 
  Time zero corresponds to 2021 May 14 00:39:53 UT.
  The red dashed line marks the average count rate obtained from the full EPIC-pn light curve.
  Bottom: Zoom-in showing the X-ray burst 
  with a time binning of 4\,s. The light curves for MOS1 and MOS2 were summed up to gain statistics.
  Time zero corresponds to the burst onset (start of first bin with a significantly increased count rate).
  }
  \label{fig:EPICLC}
\end{figure}
\begin{figure}
\centering
  \resizebox{0.91\hsize}{!}{\includegraphics{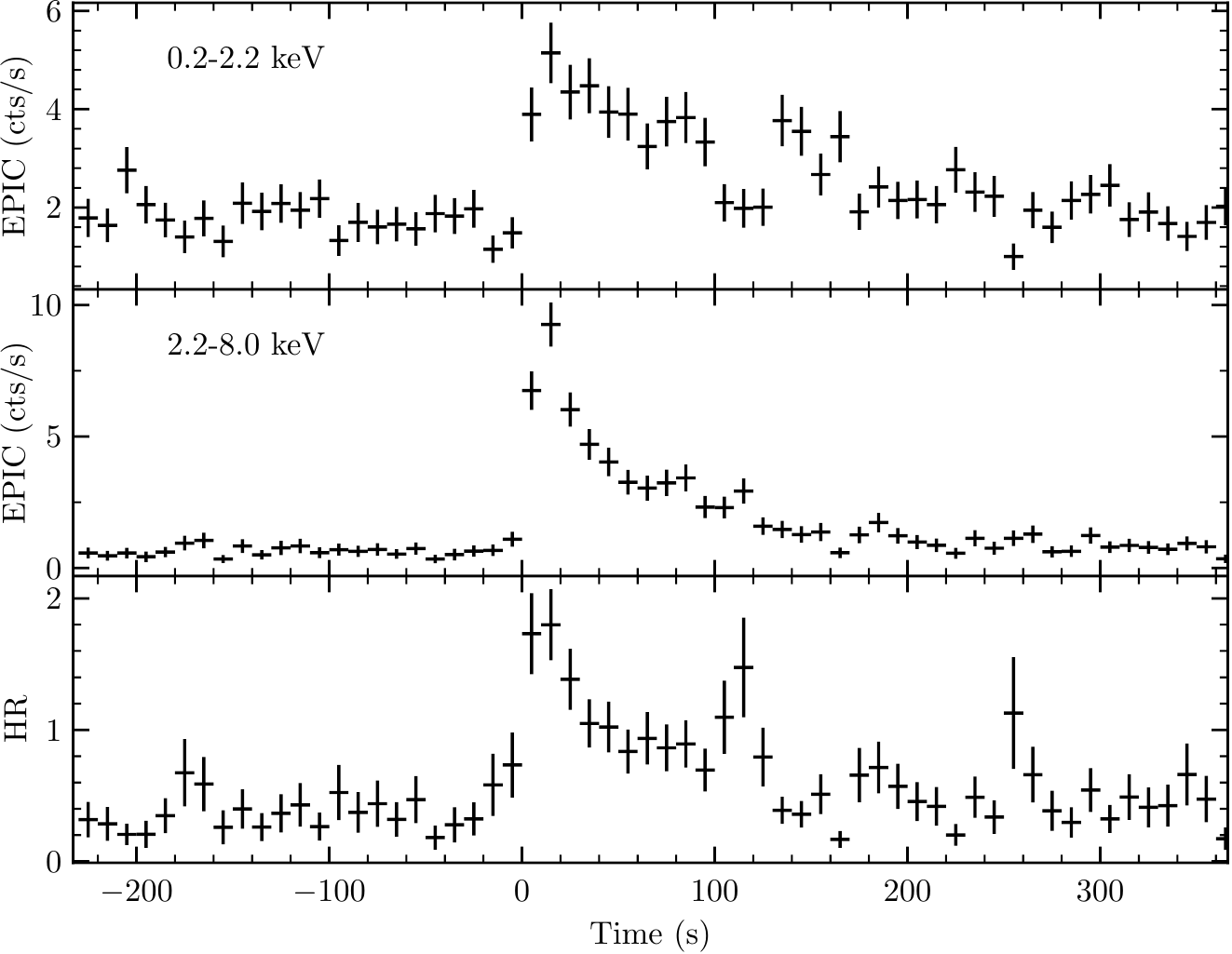}}
  \caption{Background-subtracted EPIC light curves (here MOS1, MOS2, and pn count rates are summed up) of \esrc in two energy bands with time bins of 10\,s, together with the hardness ratio (count rate in the hard band divided by the count rate in the soft band) in the bottom panel.
  Time zero corresponds to the burst onset, as in Fig.\,\ref{fig:EPICLC}
  }
  \label{fig:EPICHR}
\end{figure}

\subsection{Temporal analysis}
\label{sec:temporal}

The background-subtracted 0.2--8.0\,keV light curves of \esrc obtained from the three EPIC instruments are shown in the upper part of Fig.\,\ref{fig:EPICLC}.
A short flare-like feature is consistently seen in all three light curves, with a fast rise and exponential decay as it is demonstrated in the zoom-in in the lower part of Fig.\,\ref{fig:EPICLC}.  
A hardness ratio (HR) derived from the light curves in the soft (0.2--2.2\,keV) and hard (2.2--8.0\,keV) energy bands exhibits a similar rise and decay as the flux of the flare (Fig.\,\ref{fig:EPICHR}).
An investigation of the burst light curve (0.2--8\,keV, total EPIC count rate) at higher time resolutions down to 1\,s allowed us to estimate the rise time to $\sim$7--10\,s. The flux rise appears linear without significant dips, indicating the absence of photospheric radius expansion. Also, no significant HR variations are seen around the burst peak at 10\,s time resolution and the statistics does not allow for this to be resolved better.

The temporal and spectral evolution (seen through the HR) of the flare is consistent with that of a type-I X-ray burst seen from LMXBs.
These bursts are mainly detected from LMXBs in the Galaxy with various observatories, which allows for their spectral evolution with high time resolution to be investigated  \citep[e.g.][]{2018ApJ...856L..37K,2016MNRAS.456.4256D,2005AstL...31..681C,1987ApJ...314..266H,1977ApJ...212L..73S}. 

Apart from the burst, the EPIC light curves show a dipping behaviour during the last $\sim$1\,hr of the observation. The dips typically last 150--200\,s and reach minima, which are consistent with zero flux. The dips are more pronounced in the 0.2--2.2\,keV band, which results in increased HR. This is indicative of absorption as the origin for the dips.

\begin{figure}
\centering
  \resizebox{0.98\hsize}{!}{\includegraphics{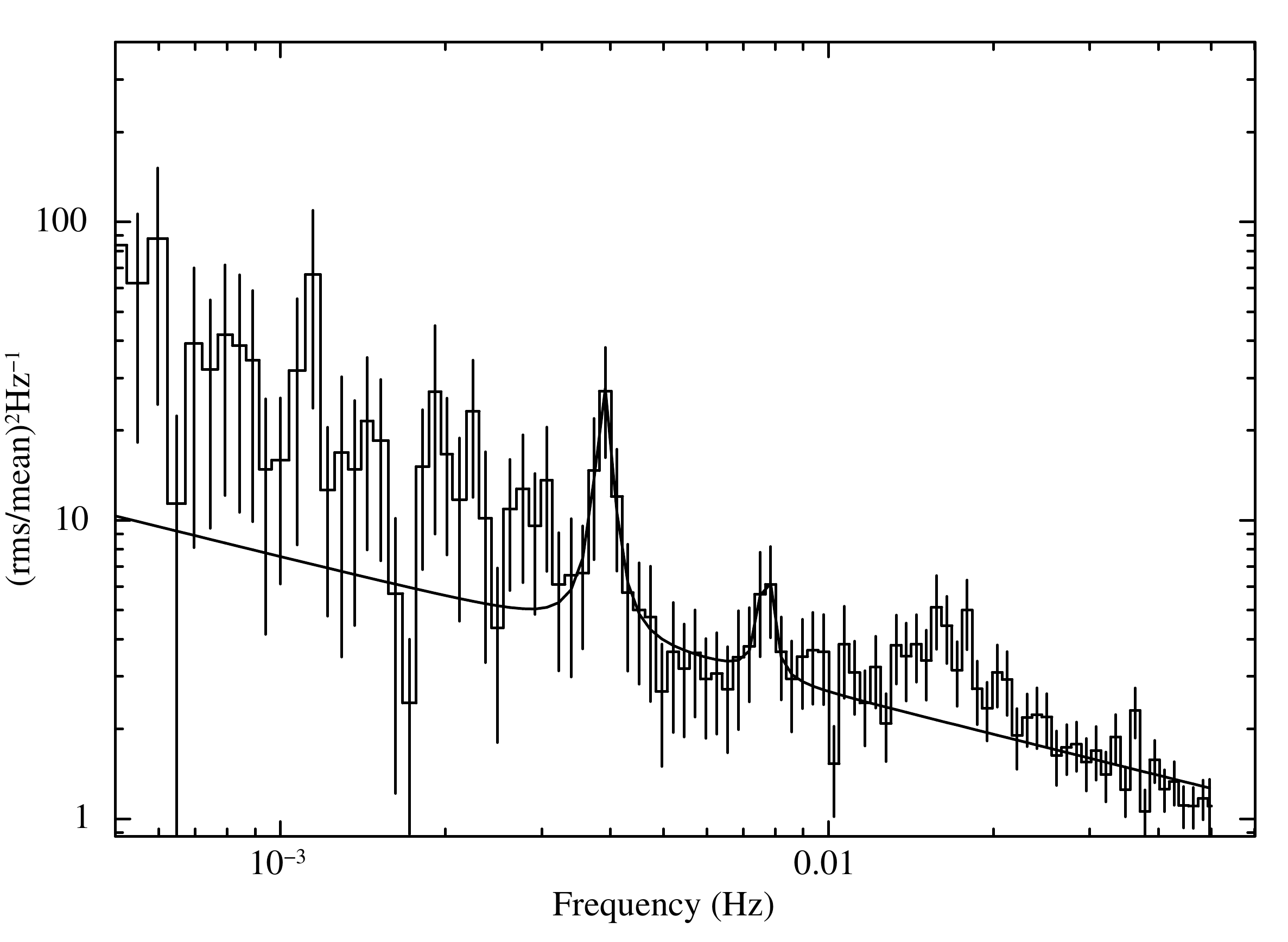}}
  \caption{PDS of \esrc produced from the EPIC-pn light curve in the energy range of 0.2--8\,keV as a histogram with error bars. The solid line represents a model fit with a power law and two Lorentzian line profiles.}
  \label{fig:EPICPDS}
\end{figure}

We created power density spectra (PDS) for both the burst and quiescent intervals from the above light curves which were divided into segments of 4096\,s with a time resolution of 10\,s. 
The PDS obtained from all these segments were rebinned with a geometric rebinning factor and averaged to produce the final PDS. 
The average PDS was also normalised such that its integral gives the squared rms fractional variability and the expected white noise level was subtracted. 
The PDS from the quiescent interval shows two putative quasi-periodic oscillations (QPOs) at $\sim$3.8\,mHz and its harmonic at $\sim$7.6\,mHz as seen in Fig.~\ref{fig:EPICPDS}. 
We fitted the PDS with a power-law component corresponding to the continuum and fitted the QPO features with two Lorentzian functions. 
The detection significance of the fundamental frequency is 2.4$\sigma$, a quality factor ($\nu$/FWHM) of $\sim$14, and an rms fractional variability of $\sim$3\% in the 0.2--8\,keV range.
The phenomenon of millihertz oscillations has been reported from a small subset of bursters at low frequencies ($\simeq$7--10\,mHz) with an rms variability of $\sim$2--3\% 
and it can be associated with the transition from stable to unstable nuclear burning, or with the regime of marginally stable nuclear burning 
\citep[see e.g.][]{2001A&A...372..138R,2007ApJ...665.1311H}.
Millihertz QPOs are observed to be suppressed after an X-ray burst 
before they reappear later \citep{2014MNRAS.445.3659L}.
Hence, we created separate PDS from the persistent emission before and after the burst. Unfortunately, the statistics in the two parts is insufficient to detect the QPOs.

\begin{figure*}
\centering
\resizebox{\hsize}{!}{\includegraphics{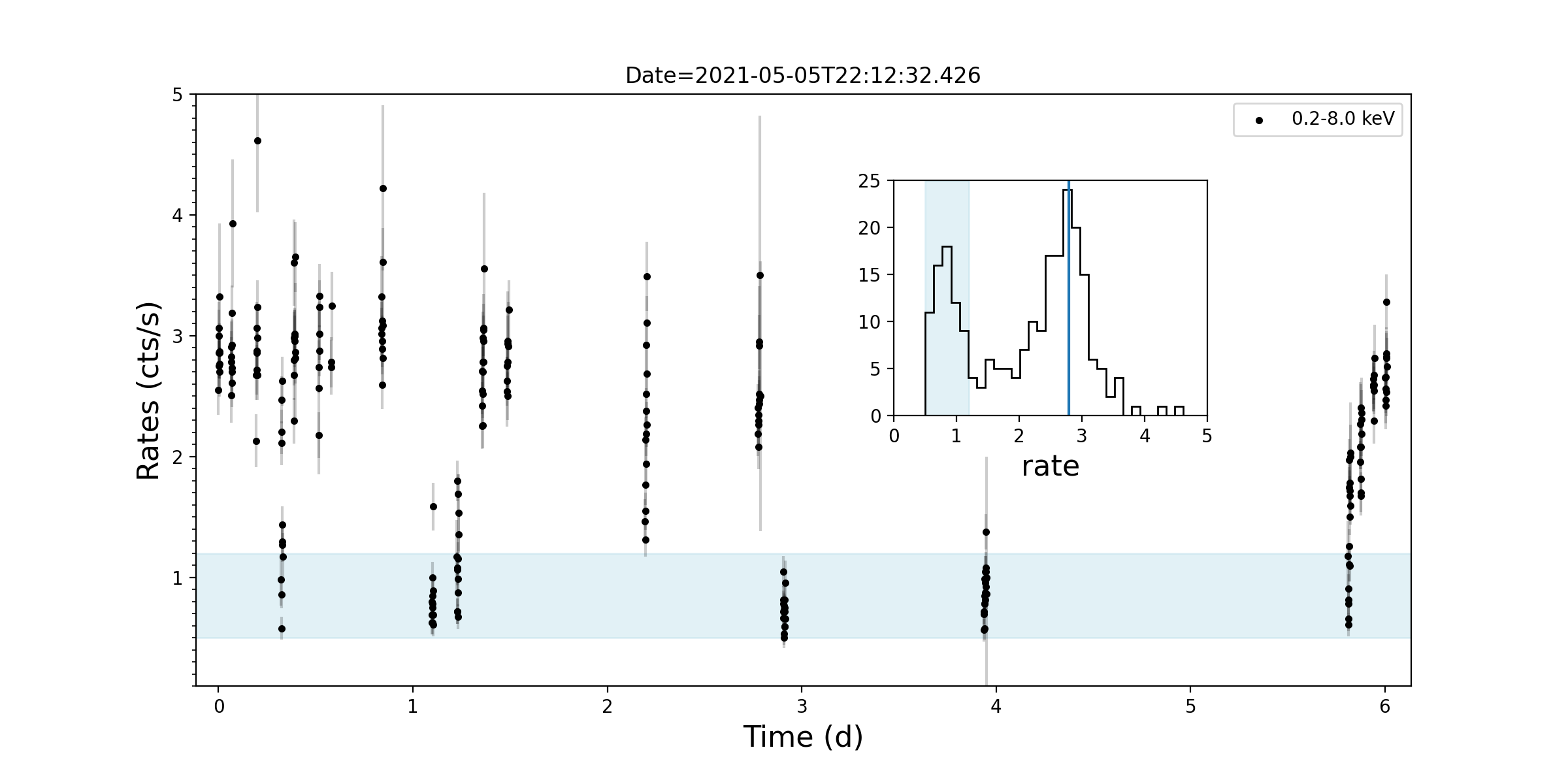}}
  \caption{\nicer X-ray light curve in the 0.2--8.0 keV energy band using 64\,s bins. 
  The distribution of fluxes (shown in the inset) is clearly bimodal, with six  low-flux states identified as shown in the shaded region (i.e. 0.5--1.2\,\cts).
  }
  \label{fig:NICERLC}
\end{figure*}

In Fig.~\ref{fig:NICERLC} we show the \nicer X-ray light curve that is used to further investigate the variability of the source. One can clearly recognise a few 
low-flux states that sporadically occur. The source seems to transition between a high and a low state as it can be seen by the histogram distribution of the count rate. The low state may be defined by the 0.5--1.2 \cts intensity range. 

\begin{figure}
\centering
  \resizebox{\hsize}{!}{\includegraphics{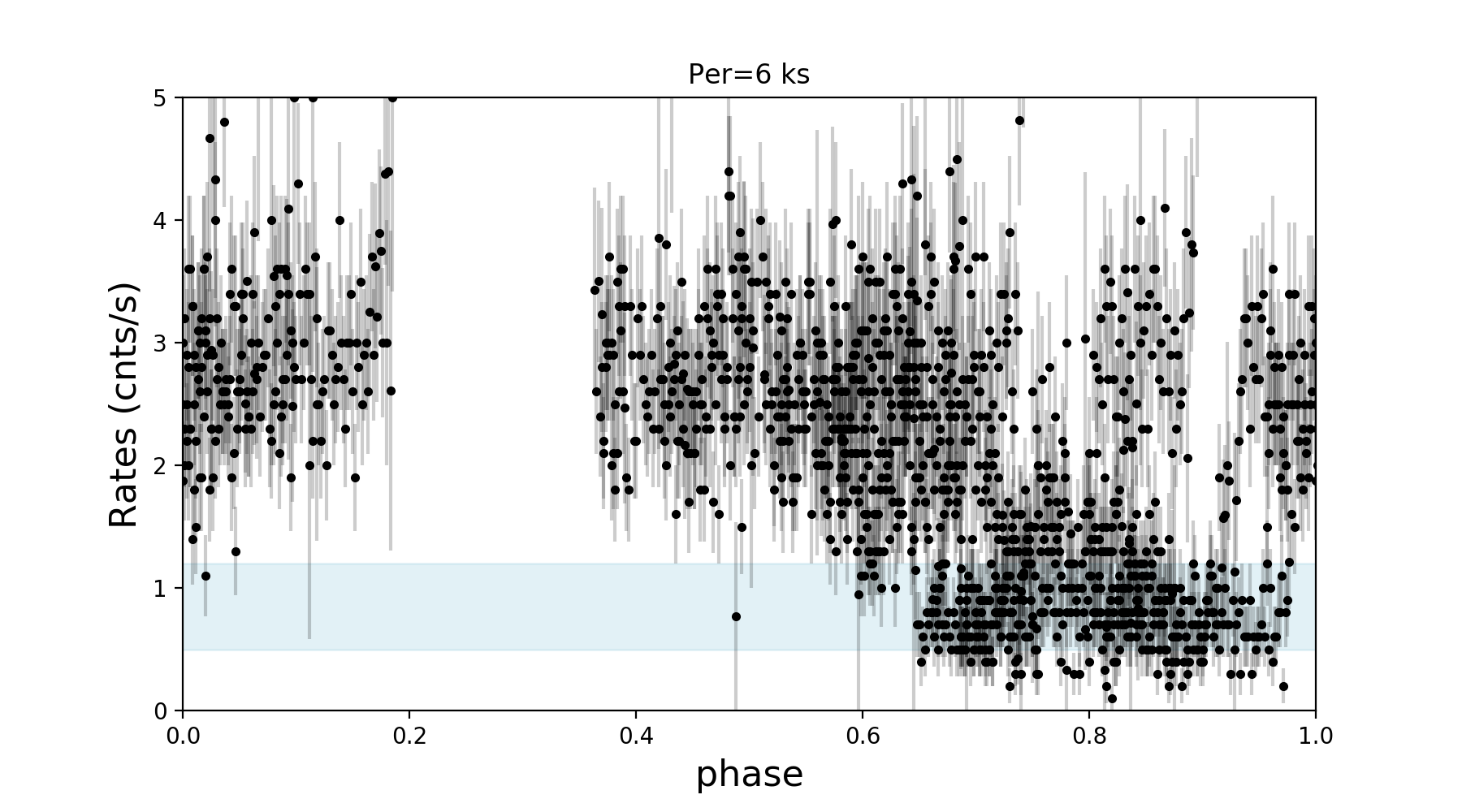}}
  \caption{\nicer light curve (10\,s bins) folded for a period of  6\,ks.
  The shaded region marks the low-flux states as in Fig.\,\ref{fig:NICERLC}.
  }
  \label{fig:NICERfold}
\end{figure}

We searched for a possible periodic nature of the dips on short ($<$20 ks) and longer (20--200\,ks) timescales.
For short timescales, we used the \nicer light curve with photons grouped into 10 s bins, and we applied a method based on the phase dispersion minimisation test described by \citet{1978ApJ...224..953S}, which has been effectively applied to light curves of other sources \citep[e.g.][]{2021ApJ...909...50V}. 
We found hints of a periodic nature at a period of $\sim$6\,ks and its harmonic. However, the low states could not be completely disentangled from the high (normal) state 
as seen in Fig. \ref{fig:NICERfold}.

\begin{figure}
\centering
  \resizebox{\hsize}{!}{\includegraphics{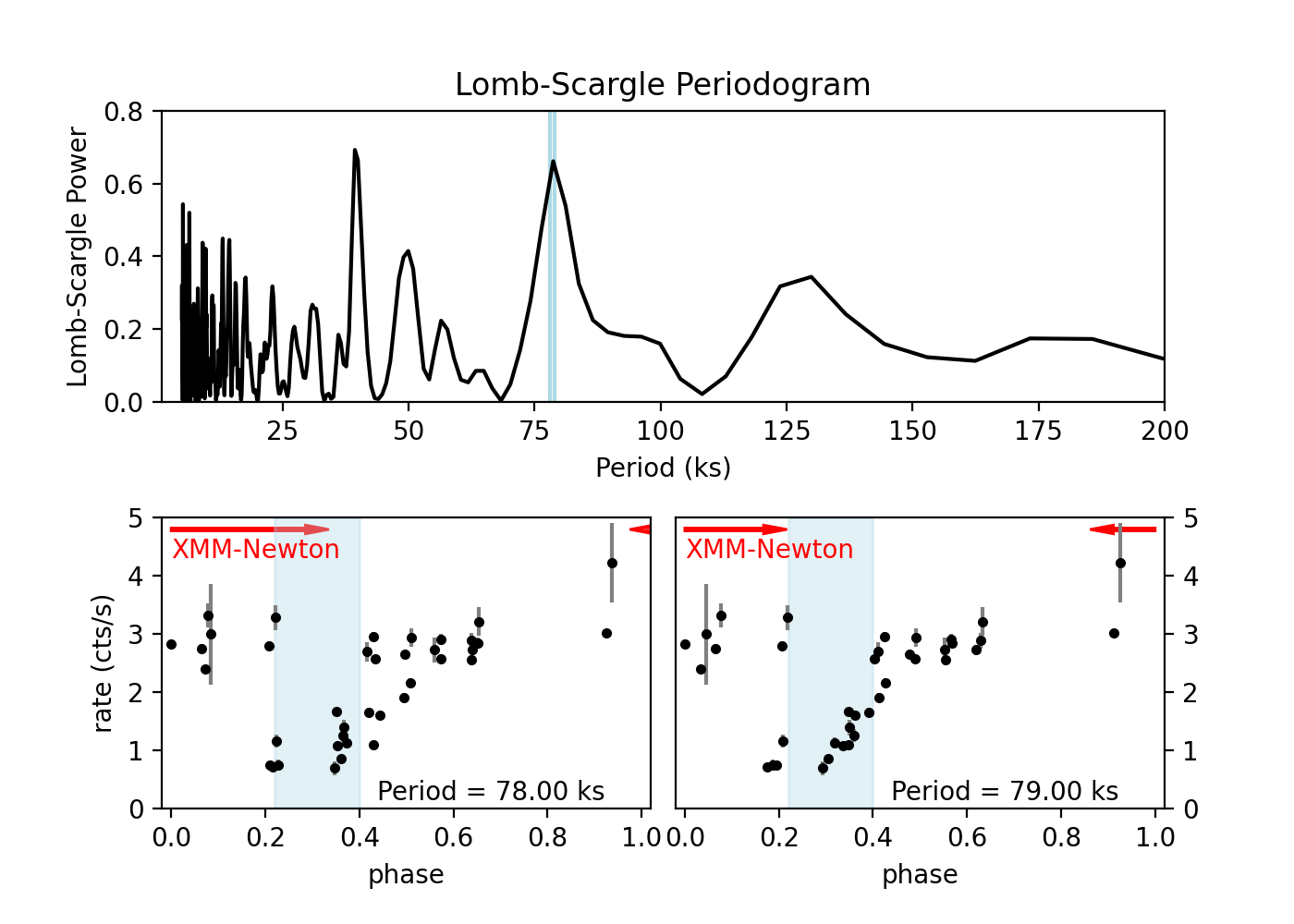}}
  \caption{
  Top: Lomb-Scargle periodogram of \nicer data, a peak is visible at a period of 79 ks.
  Bottom: \nicer light curve (300 s bin) folded for a period of 78\,ks (left) and 79\,ks (right). With red arrows, we mark the duration of the \xmm observation. The shaded region in phase outlines the low-flux states as shown in Fig.\,\ref{fig:NICERLC}.  
  }
  \label{fig:NICERfold2}
\end{figure}

Power spectra of the EPIC light curves, which cover the full frequency range between $\sim$4\expo{-5}\,Hz (limited by the observation length) and $\sim$10\,Hz (constrained by the EPIC-pn time resolution), do not show evidence for any significant signal near 6\,ks (1.7\expo{-4}\,Hz) nor at the highest frequencies.
For longer timescales, we found that the Lomb-Scargle periodogram \citep{1976Ap&SS..39..447L,1982ApJ...263..835S} was more effective. We applied the method following \citet{2018ApJS..236...16V} to the \nicer light curve. We experimented with various bin sizes between 10 s and 300 s, which yielded consistent results. The periodogram revealed peaks at $\sim$78.5\,ks and its harmonic (Fig.\,\ref{fig:NICERfold2}, top). 
We folded the \nicer light curve for periods of 78\,ks and 79\,ks as shown in the bottom of Fig.\,\ref{fig:NICERfold2}. The phases covered by the \xmm observation are indicated at the top of the plots. For the case of a period of 78\,ks, dips are expected at the end of the \xmm observation as they are observed (Fig.\,\ref{fig:EPICLC}).
In particular, from the EPIC light curves, it is evident that the associated decrease in flux does not follow the pattern expected from total or partial eclipses, but it more so resembles a dipping behaviour.

Finally, we investigated the \nicer data for possible X-ray bursts using 5\,s and 10\,s time bins, but no bursts were identified. This is not unexpected since the combined duration of the \nicer exposures is shorter than the \xmm observation.

\subsection{Spectral analysis}
\label{sec:spectral}

\begin{figure}
\centering
  \resizebox{0.95\hsize}{!}{\includegraphics{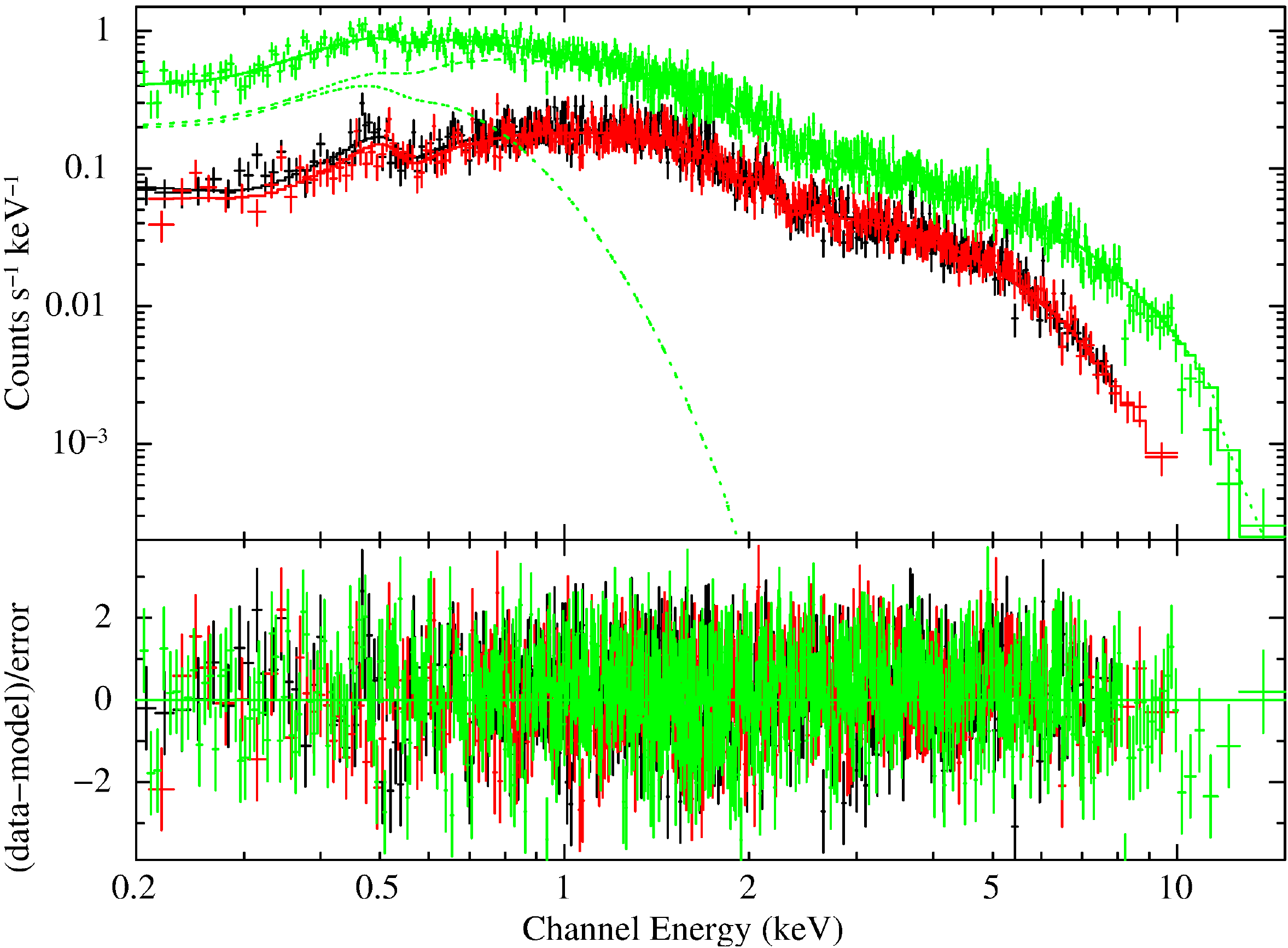}}
  \caption{
    EPIC spectra of the persistent emission from \esrc (MOS1, black; MOS2, red; and pn, green) extracted from the full observation, excluding the burst interval.
    The best-fit model consisting of a disk-blackbody and a power-law 
    component is plotted as histogram. The two emission components are indicated as dotted lines (for better visibility for the pn spectrum only). The bottom panel shows the residuals.
  }
  \label{fig:PEspectra}
\end{figure}

Analysis of the X-ray spectra was performed using \xspec v12.10.1f \citep{1996ASPC..101...17A}. 
For the spectral analysis of the burst, we extracted EPIC spectra from the quiescent emission of the full observation, excluding a $\sim$3100\,s interval around the burst (source spectra binned to a minimum of 20 counts per bin to use Gaussian statistics), and spectra from the burst starting with burst rise and exposure times of 166\,s (pn) and  174\,s (MOS) \citep[binned to a minimum of 1 count to use Cash statistics,][]{1979ApJ...228..939C}. 
The number of extracted source counts during the burst is  $\sim$1200\,counts from all three instruments together, which did not allow us to monitor spectral evolution during the burst. 
To investigate the burst spectra, we first fitted the (background-subtracted) quiescent emission using a two-component model comprised of a power law and a disk blackbody, both attenuated by photo-electric absorption along the line of sight. The spectra with the best-fit model are presented in Fig.\,\ref{fig:PEspectra} with the corresponding model parameters listed in Table\,\ref{tab:pe_spec}. 
For the fit to the burst spectra, the persistent emission components were included in the model with fixed parameters as derived above, that is to say we assume that the persistent emission did not change during the burst \citep[however, see][]{2013ApJ...772...94W,2016MNRAS.456.4256D}. For simplicity, we also adopted that the burst and persistent emission were attenuated by the same absorption.
The derived \nh value of $\sim$11.5\hcm{20} is somewhat larger than the Galactic foreground \Hone value of 8.4\hcm{20} \citep{1990ARA&A..28..215D} and consistent with 10.5\hcm{20} as derived from the HI4PI map in this direction \citep{2016A&A...594A.116H} or 13.9\hcm{20} when accounting for the contribution of \Hmol \citep{2013MNRAS.431..394W}. The latter values include the \Hone emission of the Galactic foreground and the Magellanic Bridge \citep[see also][]{2022A&A...661A..37S}.
If the absorption mainly originates in the interstellar medium, the column density determined from the X-ray spectra suggests a location of the source in the Magellanic Bridge. 

\begin{figure}
\centering
  \resizebox{0.95\hsize}{!}{\includegraphics{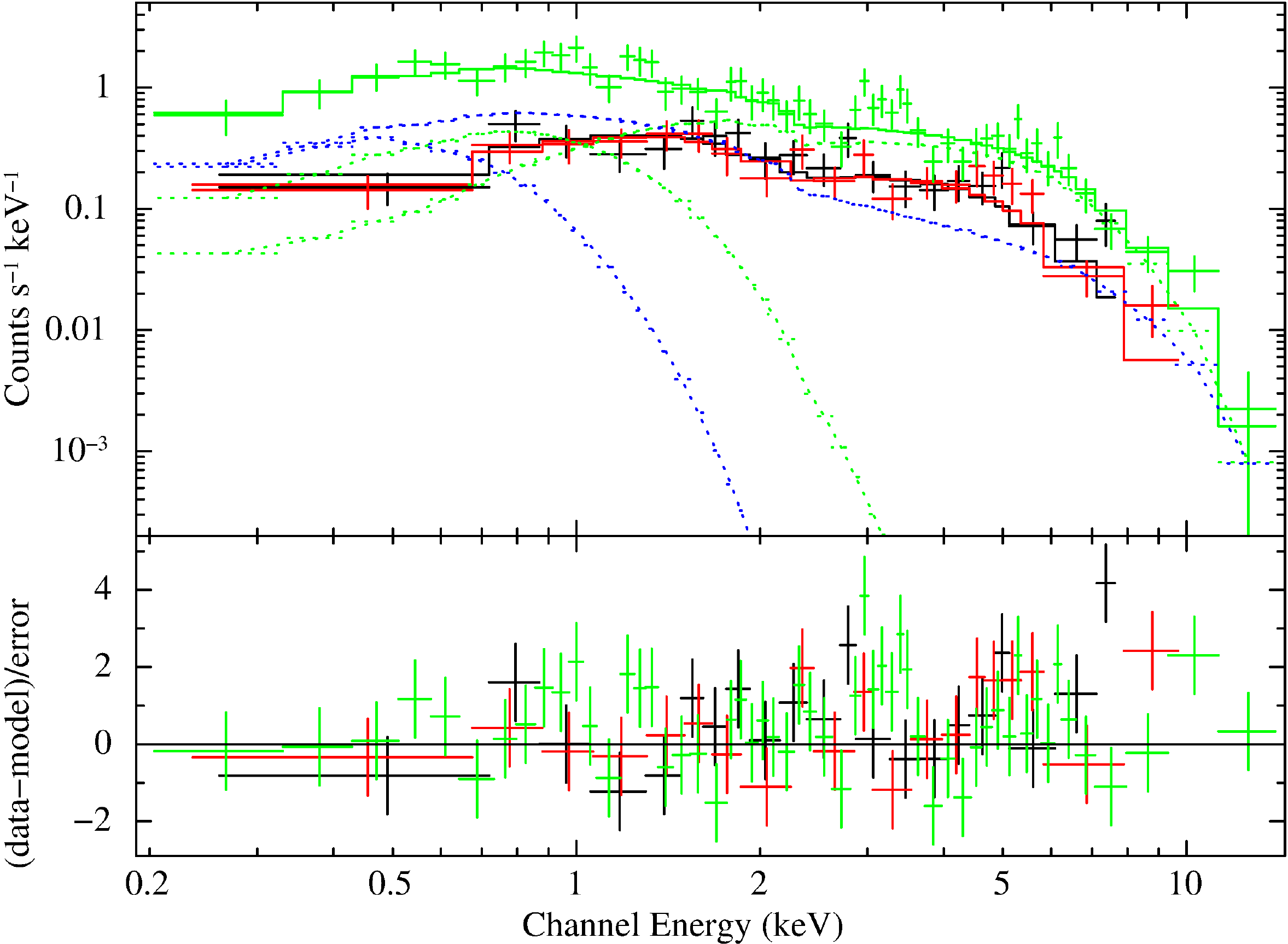}}
  \caption{
    EPIC spectra of the X-ray burst seen from \esrc (MOS1, black; MOS2, red; and pn, green).
    The best-fit model consists of two blackbody emission components in addition to the persistent emission. The best-fit model is plotted as a histogram. The two blackbody burst emission components are indicated as green dotted lines and the two components of the persistent emission as blue dotted lines (for better visibility for the pn spectrum only). The bottom panel shows the residuals.
  }
  \label{fig:BEspectra}
\end{figure}

For the burst emission, we started with a single blackbody component, which formally yielded an acceptable fit (C-statistic 758 for 951 dof), but it showed systematic residuals.
Therefore, we added a second blackbody component to account for the cooling during the burst decay, as indicated by the hardness ratio (Fig.\,\ref{fig:EPICHR}). The burst spectra with the best-fit model (C-statistic 741 for 949 dof) are presented in Fig.\,\ref{fig:BEspectra}.
For the temperatures, values of kT$_{\rm 1}$ = 1.55$^{+0.20}_{-0.16}$\,keV and kT$_{\rm 2}$ = 0.23$^{+0.11}_{-0.07}$\,keV were obtained. These two emission components dominate the spectrum at high and low energies, respectively (see Fig.\,\ref{fig:BEspectra}). The observed average burst flux in the 0.2--10\,keV band was 1.65\ergcm{-11}, which corresponds to an absorption-corrected X-ray luminosity of 6.2\ergs{36}, assuming a distance of 55 kpc (between that of the SMC and LMC). 
The bolometric luminosities inferred from the two blackbody normalisations are 6.54\ergs{36} and 0.33\ergs{36} for the high- and low-temperature component, respectively, which yield a total average bolometric burst luminosity of 6.87\ergs{36}. The average EPIC count rate over the 170\,s burst interval inferred from the light curve with a 4\,s binning (Fig.\,\ref{fig:EPICLC}) was 6.86 \cts, while the maximum reached 16.1$\pm$1.9 \cts. 
This implies a bolometric luminosity at a burst peak of at least 1.4\ergs{37}, but probably higher because the burst peak is not resolved with 4\,s
bins. The apparent emission radii for the two blackbody components, assuming spherical emission, are R$_{\rm 1}$ = 3.0$\pm$0.5\,km and R$_{\rm 2}$ = 31$^{+30}_{-16}$\,km. However, it should be noted that these can only be treated as a rough estimate because they were derived from a spectrum averaged over the burst which is expected to show large temperature variations during its time evolution.

To approximate a multi-temperature spectrum, we used a disk-blackbody model with a power-law distribution of the temperature (diskpbb in \xspec). Although this model is not physical for an application to a temperature decrease with time (which is likely more exponential), it probably allows us to better estimate the maximum temperature. In the two-blackbody model, the temperatures only provide averages over a certain time interval during the burst decay. The diskpbb model yields a similar fit quality (C-statistic of 747 for 950 dof, with one free parameter less than for the two-blackbody model) with a maximum kT = 2.9$^{+1.4}_{-0.8}$\,keV. Peak temperatures above 2\,keV are typically observed from known bursters \citep[70\% of the bursts have a peak kT$>$2\,keV,][]{2020ApJS..249...32G}.

\begin{table*}
\centering
\caption[]{Results of spectral fits to the X-ray spectra of the persistent emission.}
\begin{tabular}{l c c c c c c c @{} c @{} c}
\hline\hline\noalign{\smallskip}
\multicolumn{1}{c}{Instrument} &
\multicolumn{1}{c}{N$_{\rm H}/10^{20}$\,\tablefootmark{(a)}} &
\multicolumn{1}{c}{kT} &
\multicolumn{1}{c}{$\Gamma$} &
\multicolumn{1}{c}{C/PG-stat} &
\multicolumn{1}{c}{$\chi^2$} &
\multicolumn{1}{c}{dof} &
\multicolumn{1}{c}{F$_{\rm obs}/10^{-12}$\,\tablefootmark{(b)}} &
\multicolumn{1}{c}{L/$10^{36}$\,\tablefootmark{(c)}} &
\multicolumn{1}{c}{R$_{\rm in}\times\sqrt{cos\,\theta}$\,\tablefootmark{(d)}} \\
\multicolumn{1}{c}{} &
\multicolumn{1}{c}{(\uhcm)} &
\multicolumn{1}{c}{(eV)} &
\multicolumn{1}{c}{} &
\multicolumn{1}{c}{} &
\multicolumn{1}{c}{} &
\multicolumn{1}{c}{} &
\multicolumn{1}{c}{(\uergcm)} &
\multicolumn{1}{c}{(\uergs)} &
\multicolumn{1}{c}{(km)} \\
\noalign{\smallskip}\hline\noalign{\smallskip}
  EPIC & 11.3$\pm$1.9           & 142$^{+15}_{-12}$ & 1.42$\pm$0.03          & --   & 1404 & 1405 & 5.83$\pm$0.12          & 2.63$\pm$0.07           & 74$^{+29}_{-22}$ \\
  \noalign{\smallskip}
  \nicer & 11.7$^{+2.4}_{-2.1}$ & 129$^{+17}_{-18}$ & 1.41$^{+0.05}_{-0.05}$ &   880   & 1416 & 764   & 5.93$^{+0.2}_{-0.22}$ & 2.61$^{+0.15}_{-0.13}$  & 80$^{+50}_{-40}$ \\
  \noalign{\smallskip}
  \ero   & 11.5 fix             & 44$^{+67}_{-25}$ & 1.81$^{+0.44}_{-0.42}$  & 89.3 &  94.5 & 132 & 4.06$^{+2.04}_{-1.30}$ & 1.75$^{+5.97}_{-1.56}$  & 4260$^{+11600}_{-4130}$ \\
  \noalign{\smallskip}
  \swift/XRT & 11.5 fix             & -- & 1.64$^{+0.21}_{-0.21}$            & 86.8  &  106.6 & 120 & 5.2$^{+0.6}_{-0.7}$ & 2.3$^{+0.4}_{-0.3}$   & --  \\
\noalign{\smallskip}\hline
\end{tabular}
\tablefoot{Best-fit parameters for an absorbed two-component model consisting of disk-blackbody and power-law emission. 
Errors indicate 90\% confidence ranges.
\tablefoottext{a}{For the photo-electric absorption by gas along the line of sight, we used \texttt{tbabs} in \xspec with interstellar abundances following \citet{2000ApJ...542..914W} and atomic cross sections from \citet{1996ApJ...465..487V}.}
\tablefoottext{b}{Observed flux in the 0.2--10.0\,keV energy band.}
\tablefoottext{c}{Source-intrinsic luminosity corrected for absorption and assuming a distance of 55\,kpc (0.2--10.0\,keV).}
\tablefoottext{d}{Inner radius of the accretion disk with  inclination angle $\theta$.
                  }
}
\label{tab:pe_spec}
\end{table*}


\begin{figure}
\centering
  \resizebox{0.95\hsize}{!}{\includegraphics{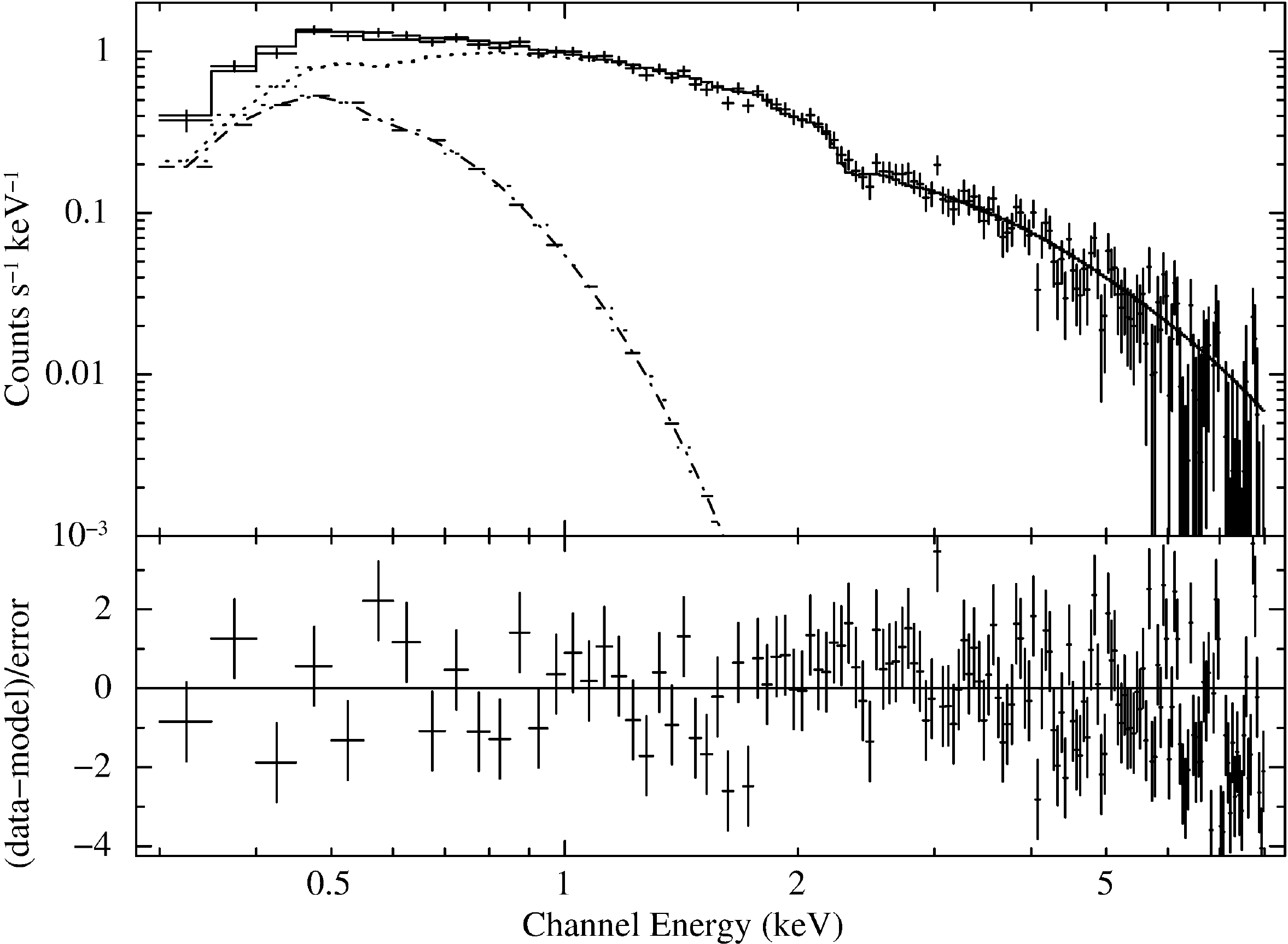}}
  \caption{\nicer spectrum of the high (normal) state (i.e. count rate $>$ 1.5 \cts). As for Fig.\,\ref{fig:PEspectra}, the contributions of disk blackbody (dash-dotted) and power law (dotted) are shown separately.
  }
  \label{fig:NICERspec}
\end{figure}

\begin{figure}
\centering
  \resizebox{0.95\hsize}{!}{\includegraphics{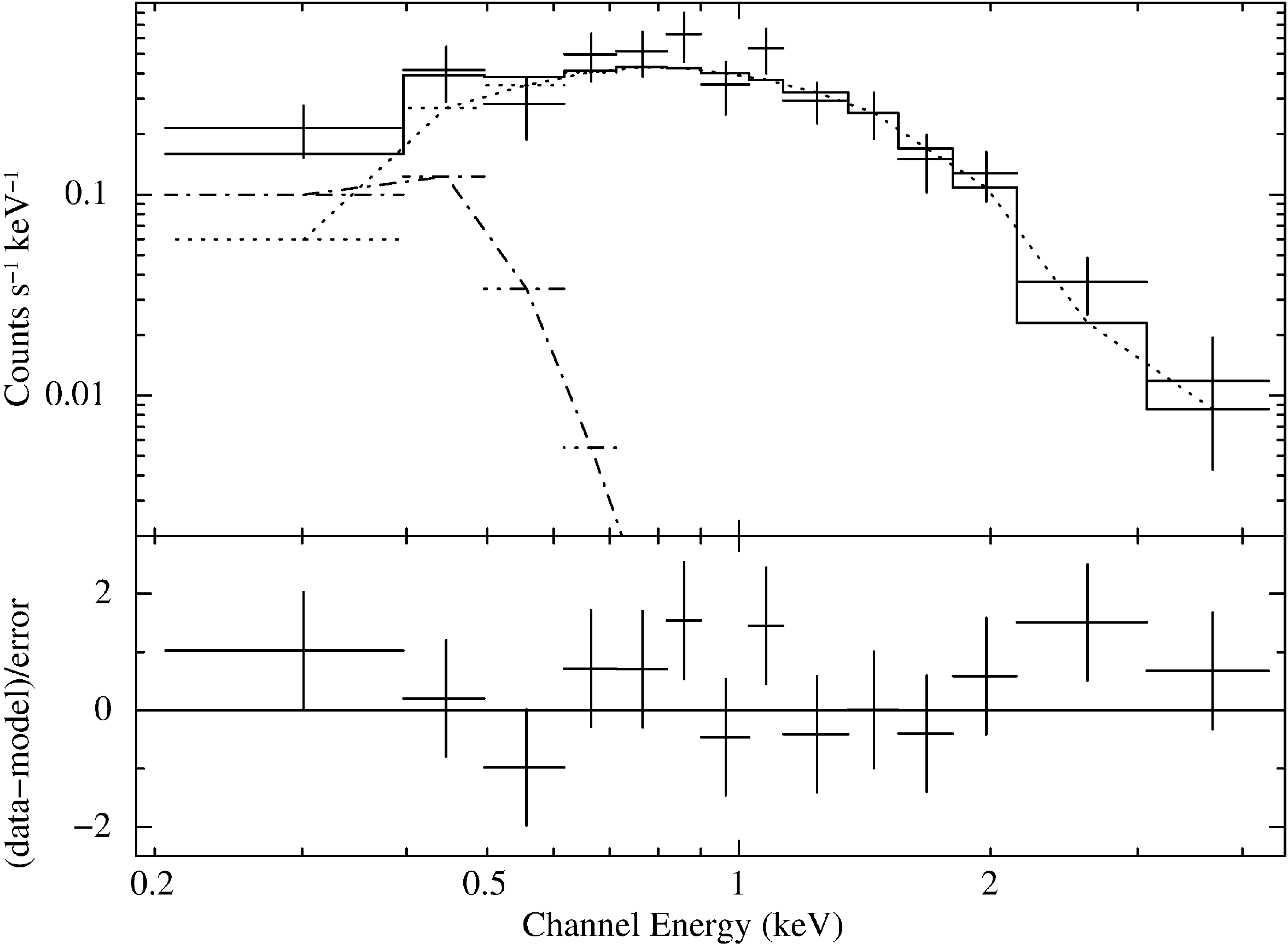}}
  \caption{
    \ero spectrum obtained from eRASS3. The best-fit model consists of a disk-blackbody (dash-dotted) and a power-law (dotted) component as used for the persistent emission seen during the \xmm observation.
  }
  \label{fig:erospec}
\end{figure}

We fitted the \nicer spectra with a similar persistent-emission model to look for possible spectral evolution.
Given the nature of the \nicer background, we performed the fitting using PG statistics within \xspec, which implements Cash statistics \citep{1979ApJ...228..939C}, with a non-Poisson background model\footnote{\url{https://heasarc.gsfc.nasa.gov/xanadu/xspec/manual/XSappendixStatistics.html}}. To maximise signal-to-noise, we only fitted \nicer data of the high (normal) state (i.e. exposure of 8\,ks).
Data were initially fitted by an absorbed power law while a soft disk blackbody was added to account for residuals at low energies. Best-fit (PG statistics of 880 for 764 dof) parameter values are given in Table \ref{tab:pe_spec}, while the best-fit model is plotted in Fig.\,\ref{fig:NICERspec}. 

We then compared the intensity of the low states seen in the \nicer light curve with the \nicer background level. For that, we extracted source and background spectra using the respective time intervals.
This resulted in a net count rate of 0.66 \cts and a background of 0.57 \cts, using the {\tt 3C50} model. Consequently, the low states are marginally brighter than the background level, and the source almost fades away during these intervals. 

We also fitted the \ero spectrum with the two-component model of a power law and disk blackbody. Because the statistical quality of the spectrum is too low to justify a fit with so many degrees of freedom, we fixed the \nh at the average value obtained from the EPIC and \nicer spectra. The spectrum with the best-fit model is presented in Fig.\,\ref{fig:erospec}. For the model parameters, readers can refer to Table\,\ref{tab:pe_spec}. The disk-blackbody temperature derived from the \ero spectrum is lower than that found from the EPIC and \nicer spectra, which may be caused by the higher low-energy contribution of the steeper power law. Fixing either the power-law index or the temperature at the EPIC/\nicer average values, however, does not bring the corresponding parameter in agreement, indicating a somewhat lower contribution of the disk component during the \ero observations. 
We note that the spectral model used by \citet{2021ATel14646....1R} includes a blackbody component instead of a disk blackbody. This has the effect that in their spectral fit, the steeper power law accounts for the soft part of the spectrum, while the hot blackbody represents the harder emission. In the model of this work, the role of the two emission components is reversed, which leads to the different best-fit parameters.

The source was detected with \swift/XRT with a net count rate of (0.094 $\pm$ 0.008) \cts. Given the small number of source counts ($\sim$150), spectra were binned to a minimum of 1 count per bin, while spectral fitting was performed based on Cash statistics.
As for the fit to the \ero spectrum, the \nh was fixed to 11.5\hcm{20}.
A simple absorbed power-law model was adequate to  explain the data (C-Statistics of 86.8 for 120 degrees of freedom, with a corresponding $\chi^2$ of 106.6, Table\,\ref{tab:pe_spec}).
The power-law photon index had a value of $\sim$1.6, consistent with what was derived from the other X-ray spectra, while the absorption-corrected X-ray luminosity was found to be $\sim$2.3\ergs{36} (0.2--10.0 keV), which is also in the range seen during the other observations.

\section{UV, optical, and near-infrared data}
\label{sec:optical}

\begin{figure*}
\centering
\resizebox{0.48\hsize}{!}{\includegraphics{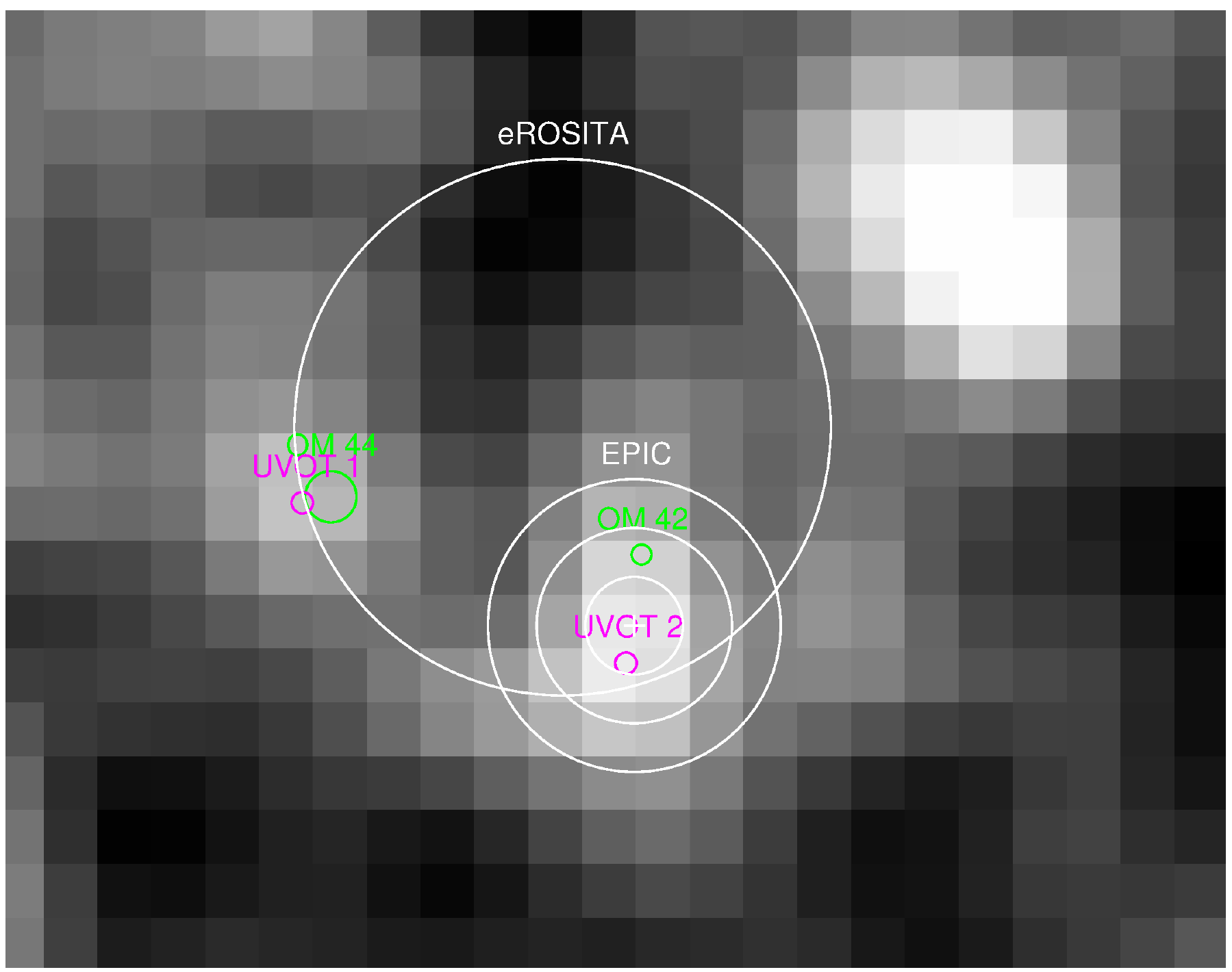}}
 \resizebox{0.48\hsize}{!}{\includegraphics{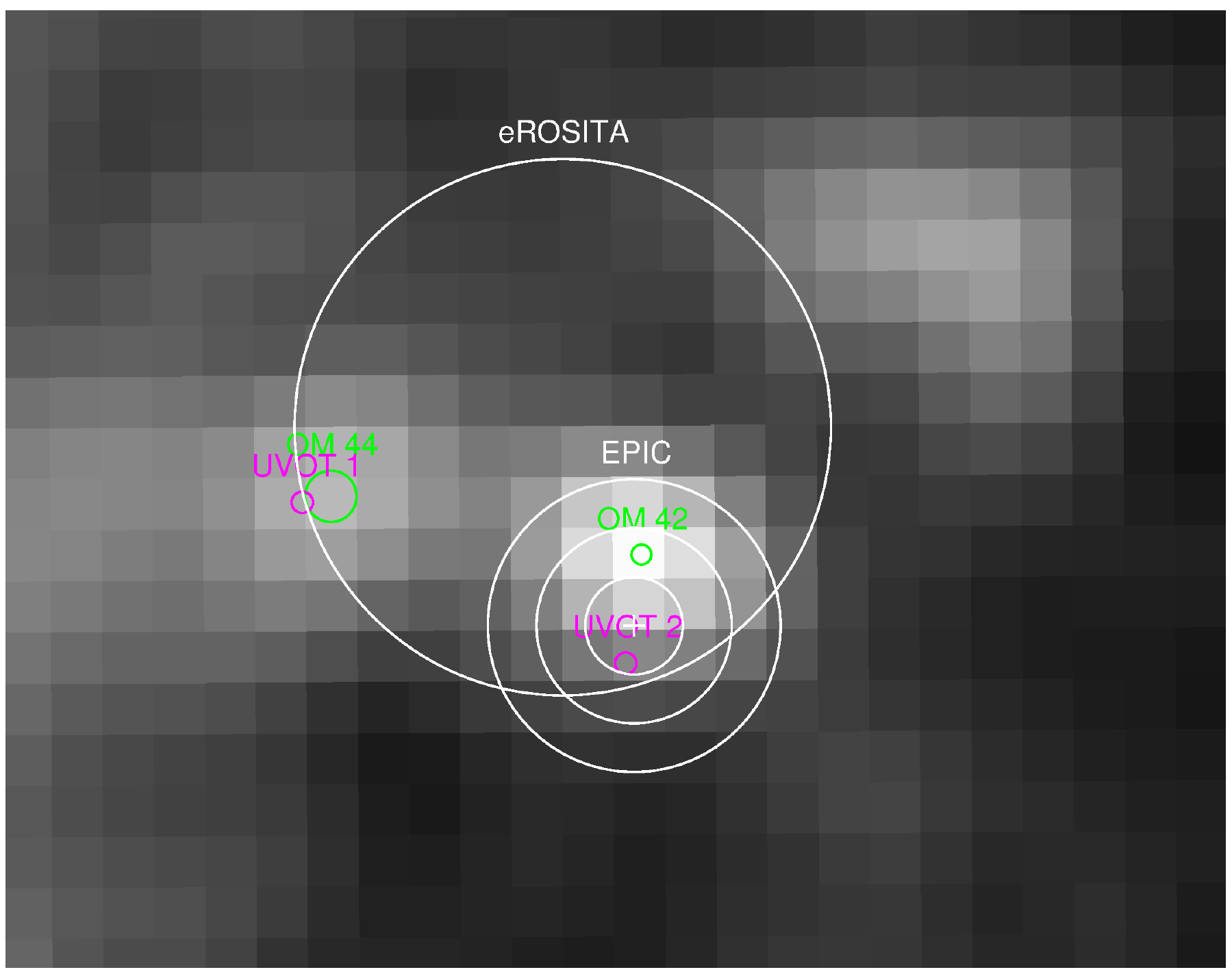}}
  \caption{UVW1-band images from \swift/UVOT (left) and \xmm/OM. 
           The large white circle (radius 5\arcsec) marks the \ero error circle from \citet{2021ATel14646....1R}, and the three concentric white circles indicate the 1, 2, and 3$\sigma$ position uncertainties obtained from the \xmm/EPIC X-ray images.
           The magenta circles mark the positions of the UVOT sources given by \citet{2021ATel14646....1R}, and the green circles (1$\sigma$ errors) indicate the sources detected in the \xmm/OM UVW1 image.}
  \label{fig:UVW1}
\end{figure*}

As described by \citet{2021ATel14646....1R}, \swift observed the region around \esrc with the Ultra-Violet/Optical telescope \citep[UVOT,][]{2005SSRv..120...95R} and the UVW1 filter\footnote{Bandwidth (full width at half maximum) of 69.3\,nm with a central wavelength of 260\,nm} with an exposure time of 1754\,s. 
We downloaded the \swift/UVOT data from the UK Swift Science Data Centre\footnote{\url{https://www.swift.ac.uk/swift_portal/}}, 
which includes the UVW1 image (shown in Fig.\,\ref{fig:UVW1}, left) and a source detection list produced by the data centre pipeline. Source magnitudes are given in the Vega system.
The two UVOT sources listed in \citet{2021ATel14646....1R} are not contained in this source list and are fainter than the faintest source (21.48\,mag) in the list, indicating their lower significance.

The source region was also observed with the \xmm optical/UV monitor telescope \citep[OM;][]{2001A&A...365L..36M}, using the UVW1 filter\footnote{Band width of 83\,nm at an effective wavelength of 291\,nm} and the detector in fast imaging mode.
The total exposure time was 20.98\,ks, which is significantly higher than the UVOT exposure.
We downloaded the pipeline products generated from the OM data from the \xmm science archive{\footnote{\url{https://www.cosmos.esa.int/web/xmm-newton/xsa}}. The OM/UVW1 image is shown in Fig.\,\ref{fig:UVW1} (right). UVOT source 1 is also detected in the OM image (source ID 44 in the OM detection list). 
Source 1 is ruled out in any case as the UV counterpart of \esrc by the improved \xmm position.
A clear source (OM 42) is detected inside the 2$\sigma$ \xmm/EPIC error radius with a UVW1 magnitude of 21.68$\pm$0.08. It is located at R.A. = 04\rahour\,05\ramin\,14\fs928 and Dec. = $-$74\degr\,52\arcmin\,00\farcs46 (J2000) with a 1$\sigma$ error of 0.19\arcsec.

Comparing OM and UVOT images (Fig.\,\ref{fig:UVW1}), UVOT source 2 might suffer from a cross-like feature, which is visible in the UVOT image. This could be caused by statistical background fluctuations due to the short exposure time or an independent transient source (at the position of UVOT\,2) blended with OM\,42.
We regard OM source 42 as more reliable as its shape is mostly circular, which is consistent with the point spread function of the instrument.
We found no counterpart at any wavelength within 5\arcsec\ to OM source 42  using the VizieR catalogue access tool{\footnote{\url{http://cdsarc.u-strasbg.fr/viz-bin/VizieR}}}.

An optical counterpart was identified in archival data from
\decam \citep[][]{2015AJ....150..150F}, a wide-field CCD camera on the Blanco 4-m telescope at the Cerro Tololo Inter-American Observatory.  Twenty-four exposures in g, r, i, and z filters from three programmes were used (including the \decam eROSITA Survey, DeROSITAS, an optical companion survey to the \ero survey).
The exposures were instrumentally calibrated by the \decam Community Pipeline \citep[][]{DCP} and catalogued for the Legacy Surveys' \citep{2019AJ....157..168D} data release (DR10 in preparation). 
The analysis uses the {\texttt Tractor} \citep{2016ascl.soft04008L} forward modelling method. The method fits models to the observations, taking the point spread functions from the exposures into account.  The {\texttt Tractor} fitting agrees with both a very compact galaxy or a point source; that is to say, the data are consistent with either a slightly extended source or a star.  The measured AB magnitudes, an absolute spectral flux density photometric system, are [$24.08\pm0.11, 23.21\pm0.07, 23.05\pm0.05, 22.27\pm0.09$] in [g, r, i, z], respectively. The photometric
calibration is tied to Pan-STARS DR1 \citep{2016arXiv161205560C} through an 
uber-cal self-calibration method \citep{2012ApJ...756..158S}. The images centred on the position obtained for the UV counterpart of \esrc are shown in Fig.\,\ref{fig:griz}.

\begin{figure*}
\centering
\resizebox{1.0\hsize}{!}{\includegraphics{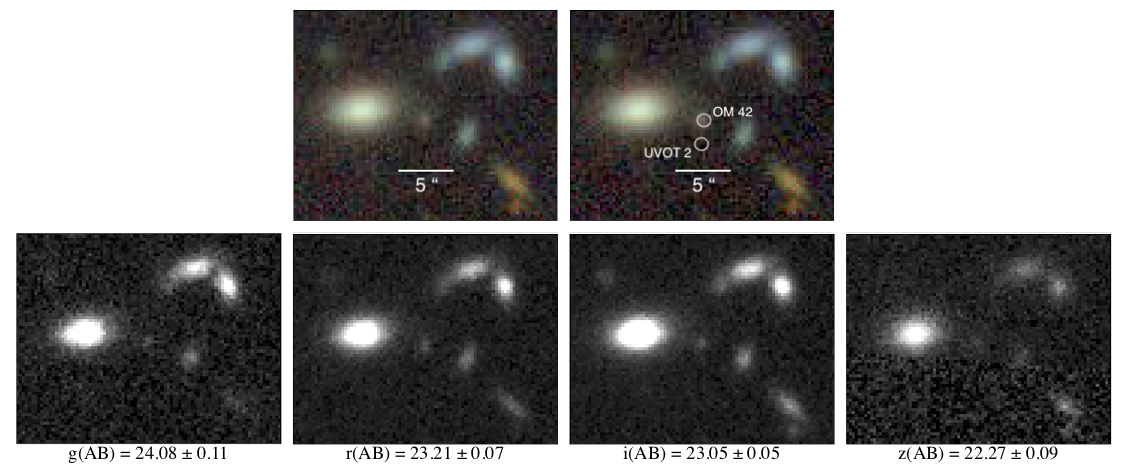}}
  \caption{Cutouts of the \decam imaging at the same size and position as Fig.~\ref{fig:UVW1}.  Each image is a coadd of many exposures in each filter and also rendered as a colour composite.  The positions of OM source 42 and UVOT source 2 are overlaid in a copy of the composite for reference.  The exposures used are public archival data included in the Legacy Surveys data release 10 (DR10).}
  \label{fig:griz}
\end{figure*}

\section{Discussion}
\label{sec:discussion}

During our \ero project to identify hard X-ray transients in the Magellanic system, an alert was generated by the \ero Near-Real Time Analysis system \citep[][]{2021A&A...647A...1P}. 
The transient, \esrc, was only detected during eRASS3 and is located in the direction of the Magellanic Bridge \citep{2021ATel14646....1R}. 

A follow-up observation with \xmm revealed an X-ray burst with energetics that suggest the source is located at the distance of the Magellanic Bridge. Assuming a distance of 55\,kpc (between that of the SMC and LMC), a burst peak bolometric luminosity of more than 1.4\ergs{37} is derived, which is typical of most X-ray bursts. Given the large distance, the statistics is low; however, we can infer certain characteristics of the burst. 

According to theory, type-I X-ray bursts, which are caused by unstable burning of the accreted matter, occur at certain mass accretion rates. Three different regimes are predicted which produce different types of X-ray bursts \citep{2000AIPC..522..359B,1981ApJ...247..267F}.

\begin{enumerate}
\item mixed H/He burning triggered by thermally unstable H ignition at low accretion rates of 
\oexpo{-14} \msun\,yr$^{-1}$  \lsim~\Mdot~\lsim~2\expo{-10} \msun\,yr$^{-1}$ resulting in a type I X-ray burst in a H-rich environment
\item pure He shell ignition after steady H burning at intermediate accretion rates of 
2\expo{-10} \msun\,yr$^{-1}$  \lsim~\Mdot~\lsim~4--11\expo{-10} \msun\,yr$^{-1}$
\item mixed H/He burning triggered by thermally unstable He ignition at high accretion rates of 
4--11\expo{-10} \msun\,yr$^{-1}$  \lsim~\Mdot~\lsim~2\expo{-8} \msun\,yr$^{-1}$
\end{enumerate}

Moreover, sedimentation of heavy elements was suggested to play an important role for the burst environment. This can lead to additional burst regimes \citep{2007ApJ...654.1022P}.

The critical accretion rates depend on metallicity, and in particular on the mass fraction of CNO 
(Z$_{\rm CNO}$ $\sim$ 0.01 for solar abundances). Between regime 1 and 2 the transition \Mdot is modified by a factor of (Z$_{\rm CNO}$/0.01)$^{1/2}$. For Magellanic Cloud metallicities between 0.2 (SMC) and 0.5 solar (LMC), the critical accretion rates are, therefore, lowered by about a factor of 0.45--0.71.

To convert the persistent luminosity of \esrc (2.6\ergs{36}) to a mass accretion rate, we assumed standard neutron star parameters (1.4\,\msun, a radius of 12 km, and a 100\% efficiency for the conversion of potential energy into X-rays). This results in 2.6\expo{-10}\,\msun\,yr$^{-1}$, which is somewhat above the critical rate between regime 1 and 2 where a He burst is expected for sub-solar metallicity.

Following \citet{2017A&A...606A.130I}, we fitted an exponential decay to the burst light curve (total EPIC count rate in the 0.2--8.0\,keV band, Fig.\,\ref{fig:EPICLC}). We find an exponential decay time of 70.6$^{+8.9}_{-7.6}$\,s (1$\sigma$ uncertainties). From the analysis of more than 1200 X-ray bursts, \citet{2017A&A...606A.130I} find that the most common decay time is 5\,s and only 1\% have decay times longer than 70\,s. For bursts driven by He burning, short ($\sim$10\,s) decay times are expected \citep{2013PrPNP..69..225P}, suggesting that the observed burst from \esrc ignited in a H-rich environment.
This is consistent with the recurrence time, which must be longer than 5.06\,hr given by the constraint of the observation start.
For He bursts, shorter recurrence times may be expected. 
For the $\alpha$ value, the ratio between integrated fluxes of persistent emission between the burst and the previous one, and the burst emission, we can infer a lower limit of 38. In the simplest model, assuming isotropic emission and all accreted matter is processed during the bursts; $\alpha$ values of $\sim$30 and $\sim$120 are expected for pure H and He burning, respectively. However, the picture is more complicated and $\alpha$ values between 10 and 1000 were observed \citep[Sect. 3.7 of][]{1993SSRv...62..223L}.

\esrc was scanned by \ero for about two days each during four epochs separated by 0.5 year intervals. The only \ero detection on May 1--2, 2021 and during follow-up observations with \nicer (May 5--12), \swift (May 12), and \xmm (May 14) suggest that the source was in a moderate outburst, lasting at least two weeks. During this outburst, the accretion rate was sufficiently high to gather enough hydrogen for the ignition of an X-ray burst. On longer timescales,  the source accretes at a much lower rate, which makes the detection of bursts in a typical X-ray follow-up observation very unlikely.

During the \nicer observations, \esrc alternated between two flux states. The high-flux level was similar to the persistent emission seen during the \xmm observation, while during low states the flux was typically a factor of 4--5 lower. 
These dips in the \nicer light curve could be due to absorption or even eclipses similar to those found in other LMXBs observed with \nicer \citep[e.g.][]{2021ATel14606....1H,2021MNRAS.503.5600B}.
The source also exhibited dipping behaviour for the last $\sim$1\,hr of the \xmm observation. The dips are resolved in the EPIC light curves, and they typically lasted for 150--200\,s and reached minimum count rates consistent with zero. During the dips, the hardness ratio increased, which suggests they are the result of absorption and obscuration by optically thick material. 
As was proposed for other dipping LMXBs, this material is likely located in a thickened region of the accretion disk where the accretion flow from the secondary star impacts on the disk \citep{2006A&A...445..179D}. In systems with sufficiently high inclination, the neutron star then gets partially occulted by this region. 
From the \nicer data, we find some evidence for a periodic nature of the dips with a period of around 78.5\,ks (21.8\,hr). Since the dips are not resolved in the \nicer light curves, the period remains uncertain, but the dip phases derived from the \xmm observation are consistent with the \nicer-derived period of 78--79\,ks. 
With an orbital period of $\sim$21.8\,hr, \esrc would be very similar to Swift\,J1858.6$-$0814 (orbital period 21.34\,hr), which also exhibits X-ray bursts and dips and, in addition, eclipses with a 1.14\,hr duration \citep{2021MNRAS.503.5600B,2022MNRAS.514.1908K}.
Eclipses could also be present in the \nicer light curves of \esrc, but the low count rate and the duration of \nicer snapshots combined with the low number of visits makes it difficult to detect the ingress and egress of eclipses and differentiate them from dips. Considering the similarities with Swift\,J1858.6$-$0814, the dipping behaviour could also be due to disk outflows \citep{2022Natur.603...52C}.

The detection of a type-I X-ray burst uniquely characterises \esrc as an LMXB with the compact object identified as a neutron star. From UV, optical, and IR images, we identified the counterpart to the X-ray source with a spectral energy distribution suggesting two emission components. The flux measurements in the g, r, i, and z bands increase with wavelength, consistent with radiation from a late-type M star (effective temperature of around 3000 K). The contribution of such a star in the UV is negligible and the bright UV emission detected by the \xmm/OM most likely originates from the accretion disk around the neutron star. An early-type OB star to account for the UV emission can definitely be ruled out as it would be  expected to be brighter than $\sim$16th magnitude in the optical B and V bands. This picture is fully consistent with the LMXB nature of the system and confirms the identification of the multi-wavelength counterpart to the X-ray source.  

The only other LMXB known in the Magellanic System is LMC\,X-2, a persistently bright system in the southern part of the LMC \citep[e.g.][]{2020MNRAS.497.3726A,2000ApJ...528..702S,1989A&A...213...97B}. The location of LMC\,X-2 is  outside the area where younger stellar populations are found, which are traced in X-rays by high-mass X-ray binaries \citep[][]{2016A&A...586A..81H} and core-collapse supernova remnants \citep[][]{2016A&A...585A.162M}. \esrc is located even further out in the Magellanic Bridge, about one-third of the angular distance to the SMC. 
The discovery of an LMXB in the Magellanic Bridge adds support to the existence of an old population \citep{2013A&A...551A..78B} and hence the theory of tidal interaction.

\begin{acknowledgements}
This work is based on data from \ero, the soft X-ray instrument aboard \srg, a joint Russian-German science mission supported by the Russian Space Agency (Roskosmos), in the interests of the Russian Academy of Sciences represented by its Space Research Institute (IKI), and the Deutsches Zentrum f{\"u}r Luft- und Raumfahrt (DLR). The \srg spacecraft was built by Lavochkin Association (NPOL) and its subcontractors, and is operated by NPOL with support from the Max Planck Institute for Extraterrestrial Physics (MPE).
The development and construction of the \ero X-ray instrument was led by MPE, with contributions from the Dr. Karl Remeis Observatory Bamberg \& ECAP (FAU Erlangen-N{\"u}rnberg), the University of Hamburg Observatory, the Leibniz Institute for Astrophysics Potsdam (AIP), and the Institute for Astronomy and Astrophysics of the University of T{\"u}bingen, with the support of DLR and the Max Planck Society. The Argelander Institute for Astronomy of the University of Bonn and the Ludwig Maximilians Universit{\"a}t Munich also participated in the science preparation for \ero.
The \ero data shown here were processed using the \eSASS software system developed by the German \ero consortium. 
This work used observations obtained with \xmm, an ESA science mission with instruments and contributions directly funded by ESA Member States and NASA. The \xmm project is supported by the DLR and the Max Planck Society.
Work on \nicer science at NRL is funded by NASA and the Office of Naval Research.
This research has made use of the VizieR catalogue access tool, CDS, Strasbourg, France (DOI: 10.26093/cds/vizier). 
The original description of the VizieR service was published in \citep{2000A&AS..143...23O}.
We thank the anonymous referee for the very useful suggestions, which helped to improve the paper.
\end{acknowledgements}

\bibliographystyle{aa} 
\bibliography{references} 

\end{document}